\documentclass[11pt]{article}

\usepackage{setspace,caption}
\usepackage{lipsum}

\usepackage[T1]{fontenc}
\usepackage{lmodern}
\usepackage[margin=1in]{geometry}
\usepackage[numbers,square]{natbib}

\usepackage{amsmath,amssymb}
\usepackage{graphicx}

\usepackage{xcolor}

\newcommand{\bra}[1]{\ensuremath{\langle#1|}}
\newcommand{\ket}[1]{\ensuremath{|#1\rangle}}
\newcommand{\braket}[2]{\ensuremath{\langle#1|#2\rangle}}
\newcommand{\expect}[1]{\ensuremath{\left\langle#1\right\rangle}}

\begin{document}

\title{Quantum phases of dipolar rotors on two-dimensional lattices}
\author{
B.P. Abolins$^a$, R.E. Zillich$^b$, K.B. Whaley$^{a,c}$\\
\textit{$^a$Department of Chemistry, University of California, Berkeley, California 94720, USA} \\
\textit{$^b$Institut f\"ur Theoretische Physik, Johannes Kepler Universit\"at Linz, Altenbergerstra\ss e 69,} \\
\textit{4040 Linz, \"Osterreich} \\
\textit{$^c$Author to whom correspondence should be addressed} \\
\\
\textit{Published version: B. P. Abolins  {\it et al.}, J. Chem. Phys. {\bf  148}, 102338 (2018)} \\ 
\textit{http://doi.org/10.1063/1.5005522}
}
\maketitle

\begin{abstract}

The quantum phase transitions of dipoles confined to the vertices of two dimensional (2D) lattices of square and triangular geometry 
is studied using path integral ground state quantum Monte Carlo (PIGS).  We analyze the phase diagram as a function of the strength of both the dipolar interaction and a transverse electric field.
The study reveals the existence of a class of orientational phases of quantum dipolar rotors whose properties are determined by
the ratios between the strength anisotropic dipole-dipole interaction, the strength of the applied transverse field,
and the rotational constant.
For the triangular lattice, the generic orientationally disordered phase found at zero and weak values of both dipolar interaction strength and applied field,
 is found to show a transition to a phase characterized by net polarization in the lattice plane as the strength of the dipole-dipole interaction is increased, independent of the strength of the applied transverse field, in addition to the expected transition to a transverse polarized phase as the electric field strength increases.
The square lattice is also found to exhibit a transition from a disordered phase to an ordered phase as the dipole-dipole interaction 
strength is increased, 
as well as the expected transition to a transverse polarized phase as the electric field strength increases. 
In contrast to the situation with a triangular lattice, on square lattices the ordered phase at high dipole-dipole interaction strength possesses a striped ordering.
The properties of these quantum dipolar rotor phases are dominated by the anisotropy of the interaction and provide useful models for developing quantum phases beyond the well-known paradigms of spin Hamiltonian models, realizing in particular a novel physical realization of a quantum rotor-like Hamiltonian that possesses an anisotropic long range interaction.
\end{abstract}

\section{Introduction}
\label{sec:intro}

In recent years quantum mechanical systems of dipolar molecules have emerged as a fascinating platform for studying a number of interesting and novel phenomena in condensed matter physics, as reviewed in
\cite{koehlerRMP06,micheliNaturePhys06,baranovPhysRep08,kremsColdMoleculesBook,hazzard:many-body_2014,Carr2009}.
Possessing a permanent electric dipole moment (in the molecule body-fixed frame),
 they interact at long range via the anisotropic dipole-dipole interaction potential
\begin{equation}
V_\mathrm{ij} = \frac{1}{4\pi\epsilon_0} \left( \frac{\mathbf{d}_i \cdot \mathbf{d}_j}{r_{ij}^3} - 3 \frac{(\mathbf{d}_i \cdot \mathbf{r}_{ij})(\mathbf{d}_j \cdot \mathbf{r}_{ij})}{r_{ij}^5} \right),
\end{equation}
where the $\mathbf{d}_i = d_i \mathbf{n}_i$ is the electric dipole moment of the $i$th molecule and $\mathbf{r}_{ij}$ is the displacement vector between the two molecules. Such interactions can be made highly tunable through the application of external fields~\cite{Buchler2007}.

Two-dimensional ensembles of 
dipolar molecules have been shown to exhibit a variety of interesting behaviors,
with a rich phase diagram even 
when the polarization is constrained to be perpendicular to the plane. Theoretical studies of fully polarized dipoles have been shown to possess a roton minimum~\cite{Mazzanti2009,Hufnagl2010,hufnaglPRL11}. They go through a series of distinct phases as the direction of the polarization is modulated~\cite{Macia2012,Macia2014}. For polarization normal to the plane of translational confinement, such systems have been shown to organize themselves into triangular lattices~\cite{Buchler2007,Astrakharchik2007}, to form a supersolid phase~\cite{Golomedov2011}, and even a unique crystalline phase that is stabilized by the zero-point motion of the dipoles~\cite{Boninsegni2013}.
Studies of fully polarized dipolar molecules confined to vertices of a lattice have been found to show 
rich phase diagrams~\cite{Goral2002,Pollet2010,Trefzger2010,Capogrosso-Sansone2010,Yamamoto2012}. In this context, the dipolar arrays are generally described using an effective Hubbard model Hamiltonian~\cite{Carr2009}, often including long-ranged corrections~\cite{Goral2002}. In particular, on triangular lattices dipolar molecules have been shown to exhibit a normal fluid phase, a Mott insulating phase, a superfluid, and a supersolid phase~\cite{Pollet2010}.
Dipolar bilayers can exhibit a pairing phase transition to pair-superfluidity\cite{maciaPRA14,filinovPRA16,astraPRA16} as well as a 
self-bound liquid state\cite{hebenstreitPRA16,raderPRA17}.

After mean field studies showed that quasi-two dimensional ensembles of harmonically confined dipoles can also exhibit roton excitations
\cite{santosPRL03,odellPRL03}, these systems have also been extensively studied. Most recently, the roton
excitations predicted by the mean field analysis have been experimentally confirmed~\cite{chomazarxiv17}.
Dipolar condensates of magnetic atoms have first been achieved with
$^{52}$Cr atoms~\cite{griesmaierPRL05,lahayeNature07}, and 
more recently with Dy~\cite{luPRL11dysprosium}
and Er~\cite{aikawaPRL12} atoms. Of particular importance for experimental realizations is an
understanding under what circumstances such systems are stable~\cite{lahayePRL08,kochNaturePhys08}.

The influence 
of trap geometry, dipole strength and short range repulsion on the stability has been
studied in the mean field approximation, indicating instability by a buckling of the condensate cloud~\cite{Ronen2007}.  In this regard, an exciting new development is the experimental generation of
self-bound droplets of trapped dipolar bosons~\cite{kadauNature16,Ferrier-Barbut_2016,chomazPRX16} which is driven
by this instability and which has been generally confirmed by calculations
\cite{Bisset_2015,Blakie_2016,Wachtler_2016,Kui-Tian_2016,maciaPRL16,cintiPRA17} using a
variety of methods.

The long range and anisotropic nature of the interaction poses challenges for a full theoretical analysis of the expected phases and dynamics of large numbers of such dipolar molecules, whether magnetically or optically trapped as ensembles, or individually localized at vertices of a lattice.
Many of the above-mentioned 
prior theoretical studies share a key common approximation, namely that molecules are treated as if they were perfectly oriented by application of an external field, with the effect of imperfect orientation in a finite external field being accommodated only through the use of an effective dipole moment.  
In cases where dipolar molecules are used to design effective spin Hamiltonians, only a few molecular excitation levels are typically included~\cite{hazzardPRL14}.
To date the full molecular excitation structure of ensembles of polar molecules have been studied only
in the mean field approximation~\cite{Zillich2011}.

The assumption of "perfect" orientation with an external field is a reasonable assumption at molecular densities low enough that the energy scale set by the rotational degree of freedom, $hB = \hbar^2/2I$ where $I$ is the molecular moment of inertia, is much larger than the dipole-dipole interaction energy, $d^2/(4\pi\epsilon_0\left<r_{ij}\right>^3)$.
In this regime the dipole-dipole interaction can be considered 
to be only a small perturbation and its effect on the orientation of the dipoles neglected, so that 
dipole orientation can be considered to be only a function of applied external field strength.
Most prior and current experiments with atoms and molecules in optical lattices are in this regime due to their large typical lattice spacing, around 300~nm to 1000~nm. Conventionally, in a lattice formed by counterpropagating laser beams, the lattice spacing is given by $\lambda/2$, where $\lambda$ is the wavelength of the trapping laser. However, several recent proposals have been made for synthesis of lattices with significantly smaller lattice spacings \cite{Ritt2006,romeroisartPRL13,Nascimbene2015,Lkacki2016,Perczel2017}, and experimental demonstration of a $\lambda/4$ lattice has already been made~\cite{Ritt2006}.  It is therefore timely to undertake a theoretical investigation of the effect of both the dipolar interactions and the external field on the molecular orientation for a general lattice of dipoles in which the lattice spacing is varied over a large range of values extending down to values where the dipole-dipole interaction becomes appreciable. 

We explore in the present work how, as we increase the density, the dipole interaction starts to affects the rotational degree of freedoms of dipoles arranged on the sites of two dimensional lattices.
The present analysis 
can potentially shed light also on the behavior of dipoles confined in 
one-dimensional trapping potentials
at high densities and subjected to a transverse electric field, where self-assembled lattices can form in the other two directions due to repulsive dipole-dipole interactions~\cite{Buchler2007,Astrakharchik2007}.
We have previously shown that at sufficiently high densities in one dimensional 
lattices, above a certain interaction strength, or conversely as the intermolecular spacing decreases and the molecular density increases, molecules will tend to spontaneously align with one another
~\cite{Abolins2011}. This gives rise to a fully polarized phase that reflects a breaking of the $O(3)$ symmetry of a distance-dependent quantum rotor Hamiltonian by the second, anisotropic term in the dipole-dipole interaction.
For a one-dimensional array, this  introduces an Ising-like $\mathbb{Z}_2$ symmetry along the axis of the array and the ordered phase for large $g$ is then a 2-fold degenerate end-to-end ordering of dipoles along the lattice axis.
In this work we explore the possibility of 
analogous transitions occurring in two dimensional systems of dipolar molecules at fixed lattice positions on square and triangular lattices.

In order the assess the full behavior of such systems and assess the limits of application for the assumption of perfect orientation, we study the behavior 
here of dipolar rotors confined to a 2-dimensional lattice ($xy$-plane) for two lattice geometries. The first lattice considered in the current work is the 2-dimensional triangular lattice. 
This choice is motivated by the theoretical predictions of dipolar molecules forming 
planar triangular crystalline lattices~\cite{Buchler2007,Astrakharchik2007} 
in presence of an explicitly defined external electric field perpendicular to the lattice directions. The second lattice considered is the 2-dimensional square lattice.
Both of these lattice structures, and many more, can be realized by confining dipolar molecules to the minima of 2-dimensional optical lattices of the corresponding geometry~\cite{windpassingerRPP13}.
In both cases the dipoles are described by a Hamiltonian of the form
\begin{equation}
H = \sum_{i=1}^N \frac{\mathbf{L}_i^2}{\hbar^2} - u \mathbf{n}_i \cdot \hat{\mathbf{e}}
	+ g \sum_{j < i} \left[\frac{\mathbf{n}_i \cdot \mathbf{n}_j}{r_{ij}^3}
	- 3 \frac{(\mathbf{n}_i \cdot \mathbf{r}_{ij})(\mathbf{n}_j \cdot \mathbf{r}_{ij})}{r_{ij}^5}\right], \label{eq:ham}
\end{equation}
where $\mathbf{L}_i$ is the usual quantum mechanical angular momentum of rotor $i$, $u = dE/hB$ with $\mathbf{E} = E\hat{\mathbf{e}}$ being the applied electric field, and $g = d^2/(4\pi\epsilon_0hBr_\mathrm{lat}^3)$ sets the strength of the dipole-dipole interacting relative to the rotational kinetic energy scale. Table~\ref{Table1:dipole parameters} shows the values of electric field and lattice spacing required to achieve the parameter values $u \geq 1$ and $g=1$, respectively, for a range of dipolar diatomic molecules. These values may be taken to correspond approximately to the onset of strong orientation and strong interparticle interactions, respectively.  We see that for the alkali halides, the strong interaction regime may be accessed at lattice spacings of a few tens of nm, while a number of the other species become strongly interacting at lattice spacings on the order of 10 nm. For all species shown here, the electric field strengths required to reach the strong orientation regime are readily accessible with current experimental capabilities.

\begin{table*}[ht]
\caption{Permanent dipole moments, rotational constants, electric field strength required to realize $u=1$, and lattice spacing required to realize $g=1$ for a range of dipolar diatomic molecules.}
 \label{Table1:dipole parameters}
  \centering
 \vspace{0.1in}
\begin{tabular}{ | c | c | c | c |c | c | } 
\hline
  {Molecule} & {d (Debye)} & {B (GHz)} & {E (kV/cm) at $u=1$} & {$r_{lat}$ (nm) for $g=1$} & {sources}  \\ 
\hline
KRb  & 0.57 & 1.10 & 3.80 &  3.56 & \cite{Gonzalez2017,Zuchowski2013} \\ 
LiCs & 5.46 & 6.53 & 2.37 &  8.83& \cite{Gonzalez2017} \\ 
NaCs & 4.70 & 1.74 & 0.73 &  12.42 &  \cite{Gonzalez2017} \\ 
CsI  & 11.69 & 0.71 & 0.12 &  30.70 & \cite{Honerjager1973,Story1976} \\ 
KBr & 10.60 & 2.43 & 0.46 &  19.10 &  \cite{NISTdiatomics} \\ 
SrO & 8.87 & 10.13 & 2.27 &  10.53 & \cite{NISTdiatomics} \\ 
SrF & 3.47 & 7.52 & 4.30 &  6.22 &  \cite{Gonzalez2017} \\ 
YO & 4.54 & 11.63 & 5.11 &  6.42&  \cite{Gonzalez2017} \\ 
YbF & 9.93 & 9.19 & 1.44 &  12.93 & \cite{Bethlem2003} \\ 
 \hline
\end{tabular}
\end{table*}

The Hamiltonian Eq.~(\ref{eq:ham}) is remarkably similar to the well-known quantum rotor model, of which there is no 
known physical example~\cite{Sachdev}. In particular, when the anisotropic term in the dipolar interaction is omitted Eq.~(\ref{eq:ham}) becomes equivalent to the $O(3)$ quantum rotor model in an external field~\cite{Sachdev,Dutta2001}. The results 
in this work show that when the short range spin-spin interaction characteristic of the conventional quantum rotor model is replaced by an anisotropic long range dipolar interaction, a new class of dipolar quantum rotor phases emerges.

\section{Methods}
\label{sec:methods}

To explicitly study the effects of the dipole interaction on all degrees of freedom of dipolar molecules,
we employ the path integral ground state quantum Monte Carlo (PIGS) method, 
sometimes referred to as the variational path integral Monte Carlo method~\cite{Sarsa2000}. The PIGS method is a straightforward extension of the well known finite temperature path integral Monte Carlo (PIMC) method that has been used extensively in recent years to study ground states of quantum systems of importance in chemistry and in condensed matter physics. These studies include van der Waals complexes~\cite{Sarsa2000}, low temperature condensed phases of helium~\cite{Sarsa2000,Rossi2009}, and more recently the elementary excitation spectrum~\cite{Macia2012} as well as the 
$T=0$~K phase diagram~\cite{Macia2014} of continuum systems of two dimensional fully polarized dipoles. The related reptation Monte Carlo method has been used to study the rotational and translational dynamics of small molecules embedded in $^4$He clusters~\cite{Cazzato2004}.
We have previously extended the PIGS method to a full simulation of both rotational and translational motion of ensembles of molecules~\cite{Abolins2011}.  We used this technique to study the behavior of dipolar rotors confined to one dimensional lattices, e.g.\ without translational degrees of freedom, finding a crossover from an unpolarized phase at low dipole-dipole interaction strength to polarized behavior at higher dipole-dipole interaction strength as 
mentioned above~\cite{Abolins2011}.
In the present study we extend this work to 2-dimensional lattices at unit filling, still fixing the translational
coordinates to the sites of a lattice.  In the following we briefly summarize the computational approach for the most general
case that both rotational and translational degrees of freedom are allowed to fluctuate.

\subsection{Path Integral Ground State for Rotating and Translating Dipolar Molecules}
\label{subsec:PIGS}

The PIGS method is quite general, being broadly applicable to the study of the ground states of arbitrary bosonic systems. PIGS belongs to the broader family of projector Monte Carlo methods which begin with a trial state or wave function. This state can formally be written in terms of the eigenstates of the Hamiltonian of interest
\begin{equation}
\ket{\Psi_\mathrm{trial}} = \sum_k^\infty c_k \ket{\Phi_k}
\end{equation}
where $H\ket{\Phi_k} = E_k\ket{\Phi_k}$ are the eigenstates of $H$. By propagating this state in imaginary time for a duration $\beta/2$,
\begin{equation}
G(\beta/2)\ket{\Psi_\mathrm{trial}} = \sum_k c_k e^{-\beta E_k/2\hbar} \ket{\Phi_k},
\end{equation}
the trial state will asymptotically approach the ground state,
\begin{equation}
\lim_{\beta \rightarrow \infty} \frac{G(\beta/2) \ket{\Psi_\mathrm{trial}}}{\sqrt{\bra{\Psi_\mathrm{trial}}G(\beta)\ket{\Psi_\mathrm{trial}}}} = \ket{\Phi_0} \label{eq:gs_exact}
\end{equation}
assuming that $c_0 \neq 0$, where 
$G(\beta/2) = e^{-\beta H /2\hbar}$ is the usual imaginary time evolution operator, simply referred to as propagator, and $\ket{\Phi_0}$ is the exact ground state of $H$. In this limit expectation values of  
an operator $O$ can be expressed as
\begin{equation}
\expect{O} = \lim_{\beta \rightarrow \infty} \frac{\bra{\Psi_\mathrm{trial}}G(\beta / 2) O G(\beta / 2)\ket{\Psi_\mathrm{trial}}}{\bra{\Psi_\mathrm{trial}}G(\beta)\ket{\Psi_\mathrm{trial}}} = \frac{\bra{\Phi_0}O\ket{\Phi_0}}{\braket{\Phi_0}{\Phi_0}}.
\label{eq:expect_exact}
\end{equation}

Breaking up the propagation into many smaller steps,
\begin{equation}
G(\beta) = [G(\tau)]^M,
\end{equation}
where $\tau = \beta / M$, suggests the use of short time approximations to the propagator, such as the well known fourth-order Trotter-Suzuki propagator~\cite{Suzuki1995} and other related approximations~\cite{Chin1997}. In what follows we used a sixth-order ``any-order'' propagator~\cite{Zillich2010} of the form
\begin{equation}
G_{2n}(\tau) = \sum_{i=1}^n c_i \left(G_2(\tau/k_i)\right)^{k_i} = G(\tau) + O(\tau^{2n+1}),
\label{eq:any_order}
\end{equation}
where $k_i = \{1, 2, 4\}$. Here, $G_2(\tau)$ is the second order propagator approximation (so-called primitive approximation) given by
\begin{equation}
G_2(\tau) = e^{-\tau V/(2\hbar)} e^{-\tau T/\hbar} e^{-\tau V/(2 \hbar)},
\end{equation}
with $H = T + V$ and $T$ is the kinetic energy and $V$ is the potential energy.

Working in a representation with coordinates $\mathbf{X}$ (e.g. in the present case
$\mathbf{X}=(\mathbf{n}_1,\dots,\mathbf{n}_N)$) and assuming a sufficiently large number
$M$ of sufficiently small imaginary propagation time steps, one arrives at an approximate expression for the expectation value
\begin{equation}
\expect{O} \approx \frac{1}{N(\beta, M)} \int d^M\mathbf{X} \,\,
	\left[\Psi_\mathrm{trial}^*(\mathbf{X}_1)
	\left(\prod_{i=1}^{M-1} G_n(\mathbf{X}_i, \mathbf{X}_{i+1}, \tau)\right)
	\Psi_\mathrm{trial}(\mathbf{X}_M)\right]
	O(\mathbf{X}_{\lfloor M/2 \rfloor + 1}),l
	\label{eq:expectO}
\end{equation}
where the integral is taken over all of the coordinates of the system, $\{ \mathbf{X}_1, \dots, \mathbf{X}_M \}$, $G_n(\mathbf{X}, \mathbf{X}', \tau) = \bra{\mathbf{X}} G_n(\tau) \ket{\mathbf{X}'}$, and $N(\beta, M)$ is a normalization constant. This form suggests the use of Monte Carlo integral evaluation of the high dimensional integral. This can be done using the well-known Metropolis algorithm~\cite{Metropolis1953} where the weight for a given path 
through the muti-dimensional configuration space of integral Eq.~(\ref{eq:expectO}) is
\begin{equation}
W(\mathbf{X}_1, \dots, \mathbf{X}_M; \beta, M) =
\Psi_\mathrm{trial}^*(\mathbf{X}_1)
\left(\prod_{i=1}^{M-1} G_n(\mathbf{X}_i, \mathbf{X}_{i+1}, \tau)\right)
\Psi_\mathrm{trial}(\mathbf{X}_M)/N(\beta, M).
\label{eq:pdf}
\end{equation}
Since the Metropolis algorithm only depends on ratios of the weights for different configurations, the normalization $N(\beta, M)$ is of no consequence and need not be evaluated.

From this we see that PIGS has many desirable qualities, namely that it can be applied to any system where the weights $W(\mathbf{X}_1, \dots, \mathbf{X}_M; \beta, M) \geq 0$, and so is generally applicable to the ground state of bosonic 
or distinguishable quantum systems. The only inputs are the system Hamiltonian, which enters through the expression for the effective propagator, and the trial wave function. This trial wave function can be as sophisticated or as simple as is desired
to balance the tradeoff between computational efficiency and complexity of evaluation of the integrand. In many situations even a constant trial wave function can be used without incurring too great a penalty in terms of efficiency~\cite{Rossi2009,rotaPRE10,Abolins2011}.
Unlike in variational Monte Carlo, the employed trial wave function does not bias the results, provided it has non-zero overlap with the ground state and the extrapolation to infinite path length is performed.
All of the approximations 
made in implementing Eqs.~(\ref{eq:expectO}) - (\ref{eq:pdf}) are in principle controllable through extrapolation to the infinite path length limit, $\beta\to\infty$, and the zero time step limit, $\tau\to 0$. These properties make the PIGS method extremely useful for studying the ground state behavior of bosonic many-body systems, although we show in the Appendix that the
convergence of our results with increasing $\beta$ is problematic close to a quantum phase transition.
While the diffusion Monte Carlo (DMC) method usually give the ground state energy with a smaller
variance, obtaining estimators that are not biased by the trial functions is not straightforward
for expectations values of operators that do not commute with $H$ \cite{casulleras95}.

To sample 
the imaginary time paths we utilized the multi-level bisection algorithm~\cite{Ceperley1995}.  We sample the orientations of the rotors according to the procedure described in~\cite{Abolins2011},
 utilizing the rotational kinetic energy propagator~\cite{Cui1997,Marx1999}
\begin{equation}
G_0(\mathbf{X}, \mathbf{X}', \tau) = \bra{\mathbf{X}}e^{-\tau T/\hbar}\ket{\mathbf{X}'} =  \prod_{i=1}^N \sum_{l=0}^\infty \frac{2l+1}{4\pi} P_l(\mathbf{n}_i\cdot\mathbf{n}_i') e^{-2\pi\tau B l(l + 1)},
\label{eq:G_rot}
\end{equation}
where $\mathbf{n}_i$ is the orientation of the $i$th molecule and $P_l(x)$ is the Legendre polynomial of degree $l$.
For computational efficiency, Eq.~(\ref{eq:G_rot}) is tabulated on a grid of values of $\mathbf{n}_i \cdot \mathbf{n}_i'$ at the beginning of a simulation and then values are calculated using linear interpolation of the values on the pre-evaluated grid throughout the course of the simulations.

\subsection{Trial Functions}
\label{subsec:trialfunctions}

In the present work we employ a Hartree trial wave function of the form
\begin{equation}
\Psi_\text{trial}(\mathbf{X}) = \prod_{i=1}^N e^{\alpha \cos \theta_i}, \label{eq:approx_wf}
\end{equation}
where $\cos \theta_i = \mathbf{n}_i \cdot \hat{\mathbf{e}}$, with a variational parameter $\alpha$ optimized for $g = 0$ and the relevant value of $u$ for each simulation. The form of the wave function in Eq.~(\ref{eq:approx_wf}) is qualitatively similar to that of a single fixed dipole in an electric field 
directed along the $z$-axis and as such is expected to capture much of the behavior 
in regions of high $u$ and low $g$. This qualitative
argument is why this particular trial wave function was employed, however it should be noted that with sufficiently long imaginary time paths it is possible to retrieve the exact behavior of the system in question even 
when using a constant trial wave function~\cite{Rossi2009,rotaPRE10,Abolins2011}. The convergence study
in the Appendix demonstrates, however, that a good trial wave function is preferable especially close
to a phase transition.

\subsection{Extended System Simulation Details}
\label{subsec:pbc}
\begin{figure}[h]!
\centering
\includegraphics[width=3in]{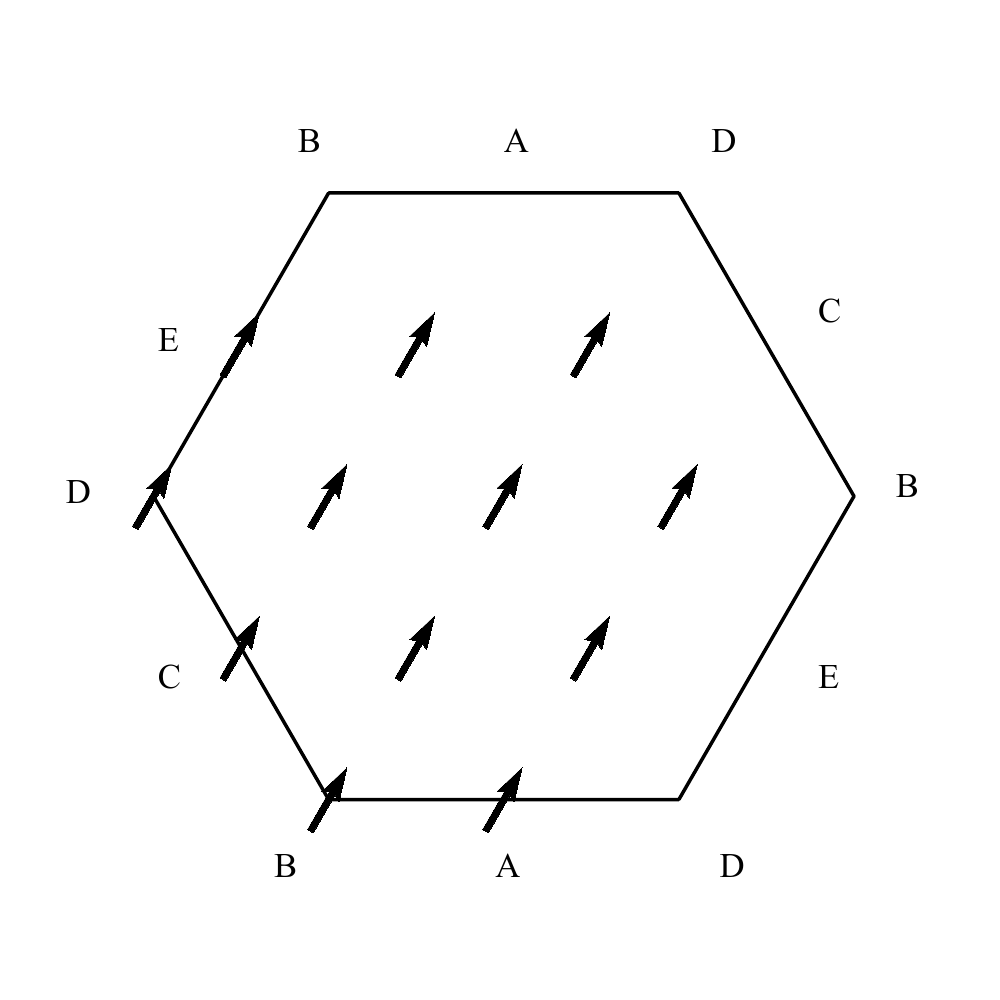}
\caption{The geometry of the hexagonal periodic simulation box. Corresponding points on the boundary are labeled with the letters A -- E.}
\label{fig:hexagonal_pbc}
\end{figure}

To describe extended systems, we employ finite sized systems with periodic boundary conditions. In two dimensions the dipole-dipole interaction, which decays with distance as $1/r^3$, requires a large cutoff to ensure that the finite sized system is representative of an extended system~\cite{Weis2003}. For our calculations on 2-dimensional lattices
the cutoff required was found to be too large to make the conventional minimum image convention~\cite{FrenkelSmit} feasible for all system sizes and so a sum over extended periodic images within a suitable chosen cutoff was employed instead, {\em i.e.} we write
\begin{equation}
H = \sum_{i=1}^N \frac{\mathbf{L}_i^2}{\hbar^2} - u \mathbf{n}_i \cdot \hat{\mathbf{e}}
	+ g \sum_{j < i} \sum_\mathbf{v} \left[\frac{\mathbf{n}_i \cdot \mathbf{n}_j}{|\mathbf{r}_{ij} + \mathbf{v}|^3}
	- 3 \frac{(\mathbf{n}_i \cdot (\mathbf{r}_{ij} + \mathbf{v}))(\mathbf{n}_j \cdot (\mathbf{r}_{ij} + \mathbf{v}))}{|\mathbf{r}_{ij} + \mathbf{v}|^5}\right], \label{eq:ham_periodic}
\end{equation}
where the sum over $\mathbf{v}$ is the sum over vectors connecting the origin of the primary simulation box to that of its periodic images~\cite{FrenkelSmit}. The long range cutoff value of $v$, yielding
$r_{max} = \max_{ij} |r_{ij} + v_{max}|$, was chosen so that $d^2/(4\pi \epsilon_0 hB r_\mathrm{max}^3) < 10^{-8}$. To properly take the anisotropic nature of the dipole-dipole interaction into account, it is also essential that the primary simulation cell has the proper symmetry.
In the case of triangular lattice simulations a hexagonal periodic simulation box was employed, illustrated in Figure~\ref{fig:hexagonal_pbc}, while for square lattice simulations a 2 dimensional square periodic simulation box was employed.

The Hamiltonian (\ref{eq:ham_periodic}) can further be written as
\begin{equation}
H = \sum_{i=1}^{N} \mathbf{L}_i^2 - u \mathbf{n}_i \cdot \hat{\mathbf{e}} + g \sum_{j<i} \mathbf{n}_i \cdot \mathbf{S}_{ij} \cdot \mathbf{n}_j \label{eq:ham_simp}
\end{equation}
with $\mathbf{S}_{ij}$ given by
\begin{equation}
  \mathbf{S}_{ij} = \sum_\mathbf{v} \left[\frac{1}{|\mathbf{r}_{ij} + \mathbf{v}|^3} \mathbb{I}
	- \frac{3}{|\mathbf{r}_{ij} + \mathbf{v}|^5} (\mathbf{r}_{ij} + \mathbf{v})(\mathbf{r}_{ij} + \mathbf{v})^\intercal\right].
\end{equation}
Because dipoles are confined to fixed points on the triangular lattice, and as a result $r_{ij}$ for all $i$ and $j$ do not change during the simulation, $\mathbf{S}_{ij}$ may be precomputed at the beginning of each simulation, leading to considerable computational savings in calculating the periodic sums in Eq.(\ref{eq:ham_periodic}).

\section{Results and Discussion}
\label{sec:results}

\subsection{Order Parameters}
\label{subsec:orderparams}

\subsubsection{Orientational phases and order parameters for triangular lattices}

\begin{figure}[h]
\centering
\includegraphics{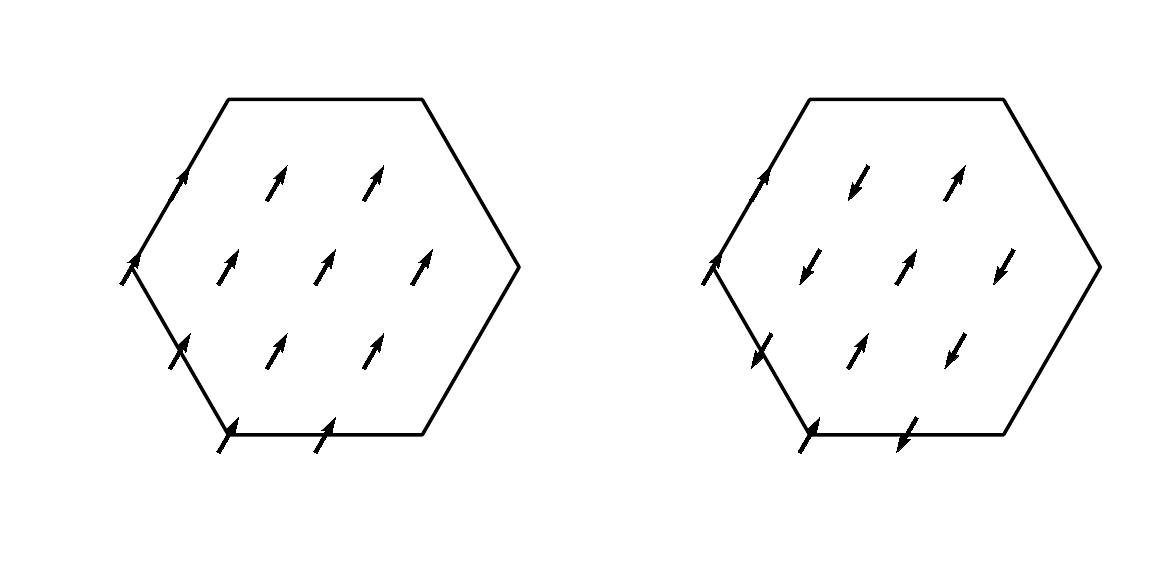}
\caption{The considered orderings in the limit $g \gg 1$ and $u \ll 1$ for a system of 12 dipoles with the primary simulation box outlined for reference. On the left is a fully polarized ordering, which is degenerate with all other configurations that are fully polarized in the lattice plane, and on the right is one of six degenerate striped orderings, two along each of three triangular lattice axes. The dipoles not pictured on the right and top boundaries correspond to particles on the opposite boundary.}
\label{fig:tri_ord}
\end{figure}

In order to assess the ordering of the ground state in the strong interaction (classical) limit, several different types of orderings were considered. In the case of triangular lattice simulations the most plausible of these are a fully polarized ordering, where all dipoles are oriented in the same direction in the plane of the lattice, and a striped ordering where alternating rows or columns of dipoles are oriented in the same direction, and all dipoles are aligned with the same axes, leading to a vanishing net polarization. These orderings are illustrated in Figure~\ref{fig:tri_ord} for a hexagonal simulation box with periodic boundary conditions.

\begin{figure}[h]
\centering
\includegraphics{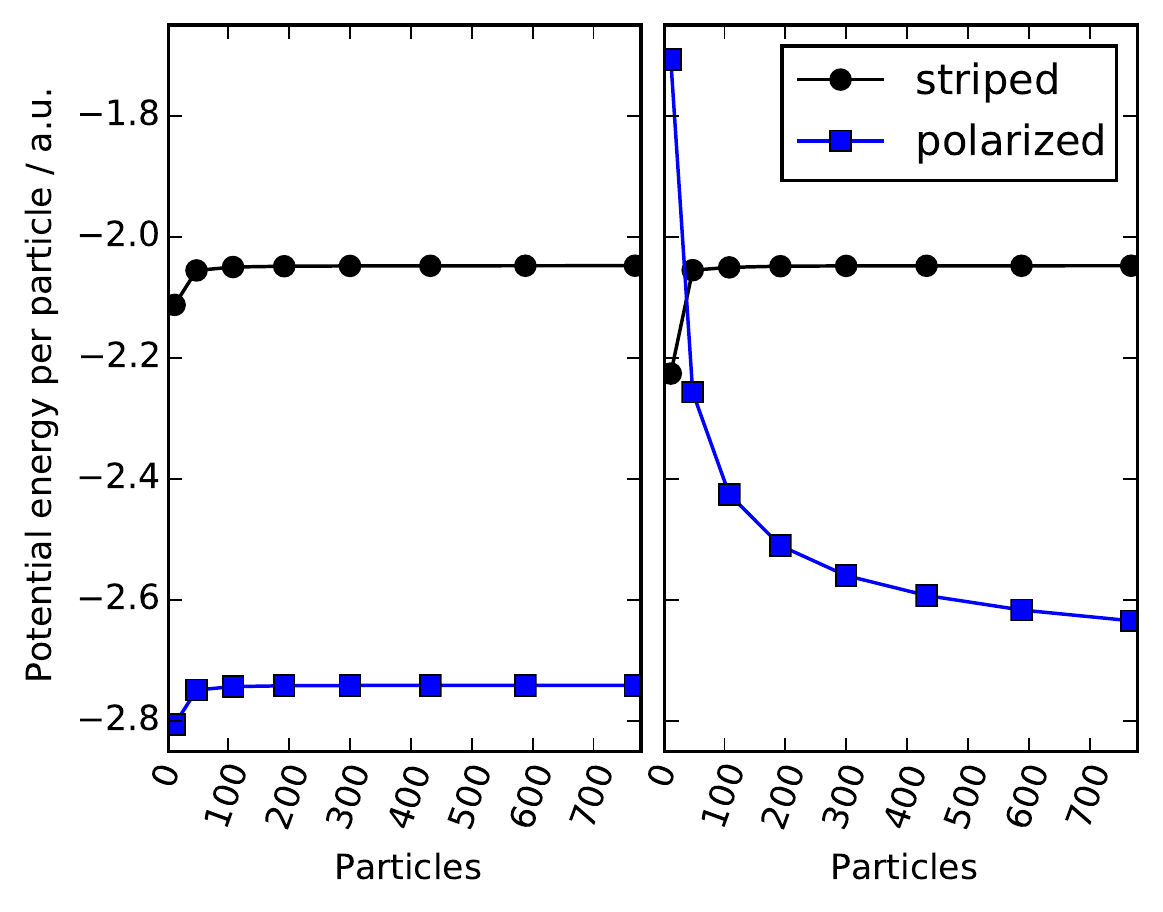}
\caption{The classical potential energy per particle of striped and polarized configurations 
in the triangular lattice as a function of system size, with $C_\text{dd}/(4\pi r_\text{lat}^3) = 1$~a.u. and $u = 0$ in a hexagonal unit cell under the periodic sum convention with a cutoff radius of $100$ lattice sites (left) and the minimum image convention (right).}
\label{fig:tri_ord_en}
\end{figure}

The correct classical ground state ordering 
of the different phases was determined by computing and comparing the potential energy per particle of each configuration as a function of the lattice size.  The resulting classical energies on the triangular lattice are depicted in Figure~\ref{fig:tri_ord_en} as function of system size. We see that the polarized ordering is by far the lower energy of the two classical ordered configurations.
This classical analysis also reveals that the energy of the fully polarized state is invariant to arbitrary rotation of the orientation of the dipoles in the lattice ($xy$-)plane, exhibiting $O(2)$ symmetry with respect to dipole orientation in this plane.  The left and right panel show the potential energy per particle calculated with the periodic sum convention explained in section~\ref{subsec:pbc} and with the conventional minimum image convention.  The comparison illustrates that the latter is severely biased by system size in the polarized case, which would require a prohibitively large simulations size.
The evaluation of the potential energy with the periodic sum convention used in this work has essentially no finite size bias, therefore our simulations can be quite small, with typically 48 dipoles in the triangular case.
Notice that for the smallest system size, the nearest image convention erroneously predicts that the striped configurations have lower potential energy than the fully polarized configurations.

With the fully polarized in-plane ordering being established as a likely candidate for the ground state in the classical strong interaction limit, we can then construct an order parameter
for the triangular lattice that is maximal in this fully polarized state. The quantity
\begin{equation}
\phi_\mathrm{pol} = \sqrt{\expect{\frac{1}{N} \sum_{i=1}^N n_i^x}^2 + \expect{\frac{1}{N} \sum_{i=1}^N n_i^y}^2}, \label{eq:tri_orderp}
\end{equation}
where $n_i^\alpha$ is the $\alpha$ component of the orientation vector, $\mathbf{n}_i$, takes on a maximum in the fully ordered state and vanishes in a state with randomly distributed dipoles.  $\phi_\mathrm{pol}$ can thus serve as an order parameter for formation of the fully polarized in-plane state during a quantum Monte Carlo simulation.  It should be noted that since this quantity is unsigned, it will not fully vanish in a simulation of a disordered state since there will always be some non-vanishing net polarization in some direction, albeit small and randomly distributed. Consequently the average value over the course of a simulation will be small but finite. In addition,  the following quantity indicative of the polarization along the applied field direction
\begin{equation}
\phi_z = \expect{\frac{1}{N} \sum_{i=1}^N n_i^z}
\end{equation}
provides an order parameter for polarization in the transverse direction.

\subsubsection{Orientational phases and order parameters for square lattices}

In the case of square lattice simulations similar 
orientational orderings were considered, with the key difference from the orderings for the triangular lattice being that the resulting polarization and striping is now  defined along the cartesian directions. Because of the symmetry of the square lattice there are only four equivalent fully polarized configurations: one in which all of the dipoles are completely polarized along the $x$-axis, and one in which the dipoles  are completely polarized along the $y$-axis. All other polarizations are higher in energy, in contrast to what is found on triangular lattices.

\begin{figure}[h]
\centering
\includegraphics{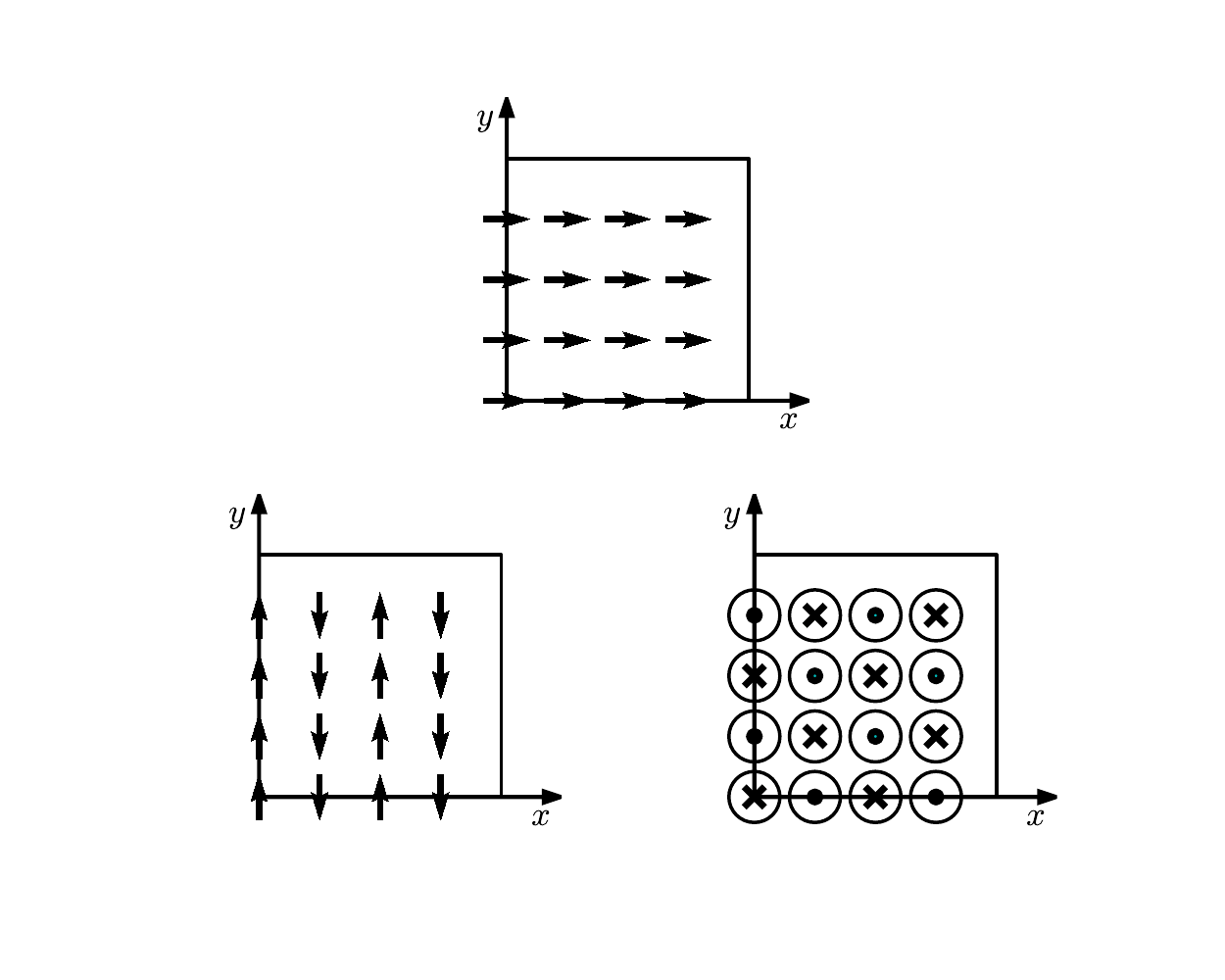}
\caption{The considered orderings in the limit $g \gg 1$ and $u \ll 1$ for a system of 16 dipoles with the primary simulation box outlined for reference. On top is a fully polarized ordering aligned along the $x$-axis, which is degenerate with the ordering in which all dipoles are aligned along the $y$-axis. On the bottom left is one of four degenerate striped orderings, two along each of two cartesian lattice axes. On the bottom right is one of two possible checkerboard orderings where an "x" denotes a dipole oriented in the negative $z$ direction and a dot denotes a dipole oriented in the positive $z$ direction.}
\label{fig:sq_ord}
\end{figure}

On square lattices the symmetry of the lattice admits another possible attractive ordering in which dipoles are aligned with the $z$-axis and where nearest neighbor dipoles are oriented anti-parallel to one another. 
We refer to this ordering as the checkerboard ordering.   Striped, fully polarized, and checkerboard orderings for the square lattice are depicted in Figure~\ref{fig:sq_ord}.

\begin{figure}[h]
\centering
\includegraphics[width=5in]{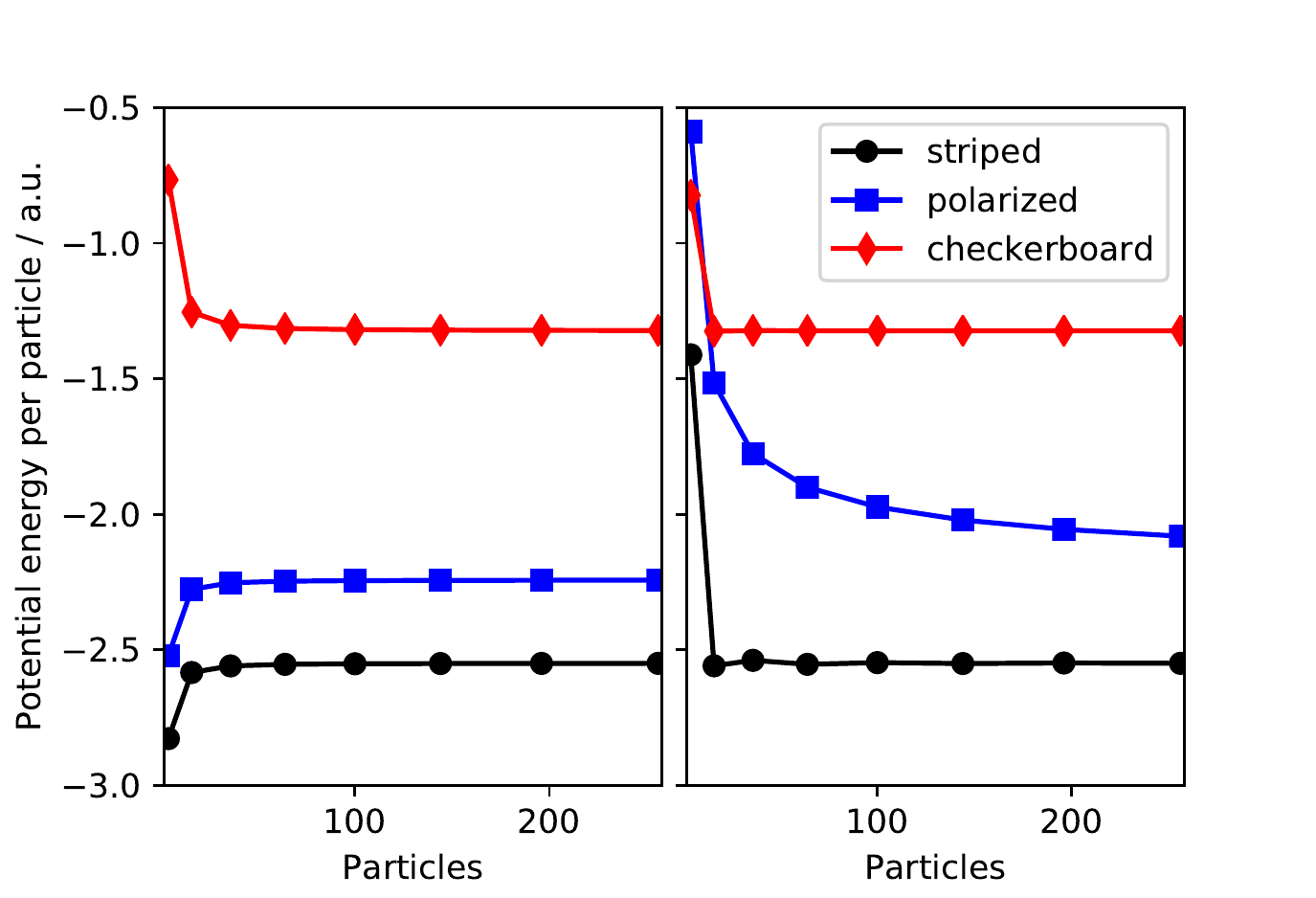}
\caption{The classical potential energy per particle of striped, polarized, and checkerboard configurations as a function of system size, with $C_\text{dd}/(4\pi r_\text{lat}^3) = 1$~a.u. and $u = 0$ on a square lattice under the periodic sum convention with a cutoff radius of $100$ lattice sites (left) and the minimum image convention (right).}
\label{fig:sq_ord_en}
\end{figure}

As in the case of triangular lattices, an analysis of the classical potential energy of various orderings was performed. In this case, however, it was found that for all lattice sizes the two striped orderings had the lowest potential energy, depicted in Figure~\ref{fig:sq_ord_en}. Again, the left and right panel show the potential energy per particle calculated with the periodic sum convention and with the conventional minimum image convention, which
again highlights the advantage of the periodic sum convention regarding finite size effects.

In order to detect the relevant striped orderings the following quantity
\begin{equation}
\phi_{xy} = \sqrt{\expect{n_{x\text{-stripe}}}^2 + \expect{n_{y\text{-stripe}}}^2}, \label{eq:sq_xystripe}
\end{equation}
where the operators
\begin{equation}
n_{x\text{-stripe}} = \frac{1}{N}\left| \sum_{i=1}^{\sqrt{N}} \sum_{j=1}^{\sqrt{N}} (-1)^i n^x_{i\sqrt{N} + j} \right| \label{eq:x_stripe}
\end{equation}
and
\begin{equation}
n_{y\text{-stripe}} = \frac{1}{N}\left| \sum_{i=1}^{\sqrt{N}} \sum_{j=1}^{\sqrt{N}} (-1)^j n^y_{i\sqrt{N} + j} \right| \label{eq:y_stripe}
\end{equation}
pick out striped configurations aligned along the $x$ and $y$ axes, respectively, takes on a maximum in the fully ordered state when $g$ is very large and vanishes when dipoles are oriented randomly. The quantity $\phi_z$ remains unchanged as a way to detect polarization in the $z$ direction. The quantity
\begin{equation}
\phi_\text{checkerboard} = \expect{\frac{1}{N}\left| \sum_{i=1}^{\sqrt{N}} \sum_{j=1}^{\sqrt{N}} (-1)^{i + j} n^z_{i\sqrt{N} + j} \right|}
\end{equation}
is maximal in the case of checkerboard ordering.
However this was not observed to any appreciable extent.

\subsection{PIGS Results for Dipoles on Triangular Lattices}
\label{subsec:results}

Systems of 48 dipoles confined to triangular lattices were simulated using the PIGS method and varying the values of $u$ and $g$ from 0 to 3 each.
We chose a time step of $\tau = 0.0375$ $(2\pi B)^{-1}$ and a imaginary time path length of $\beta = 5.1$ $(2\pi B)^{-1}$.
A detailed study of the bias introduced by finite $\tau$ and $\beta$ can be found in the Appendix.

\begin{figure}[p]
\centering
\includegraphics{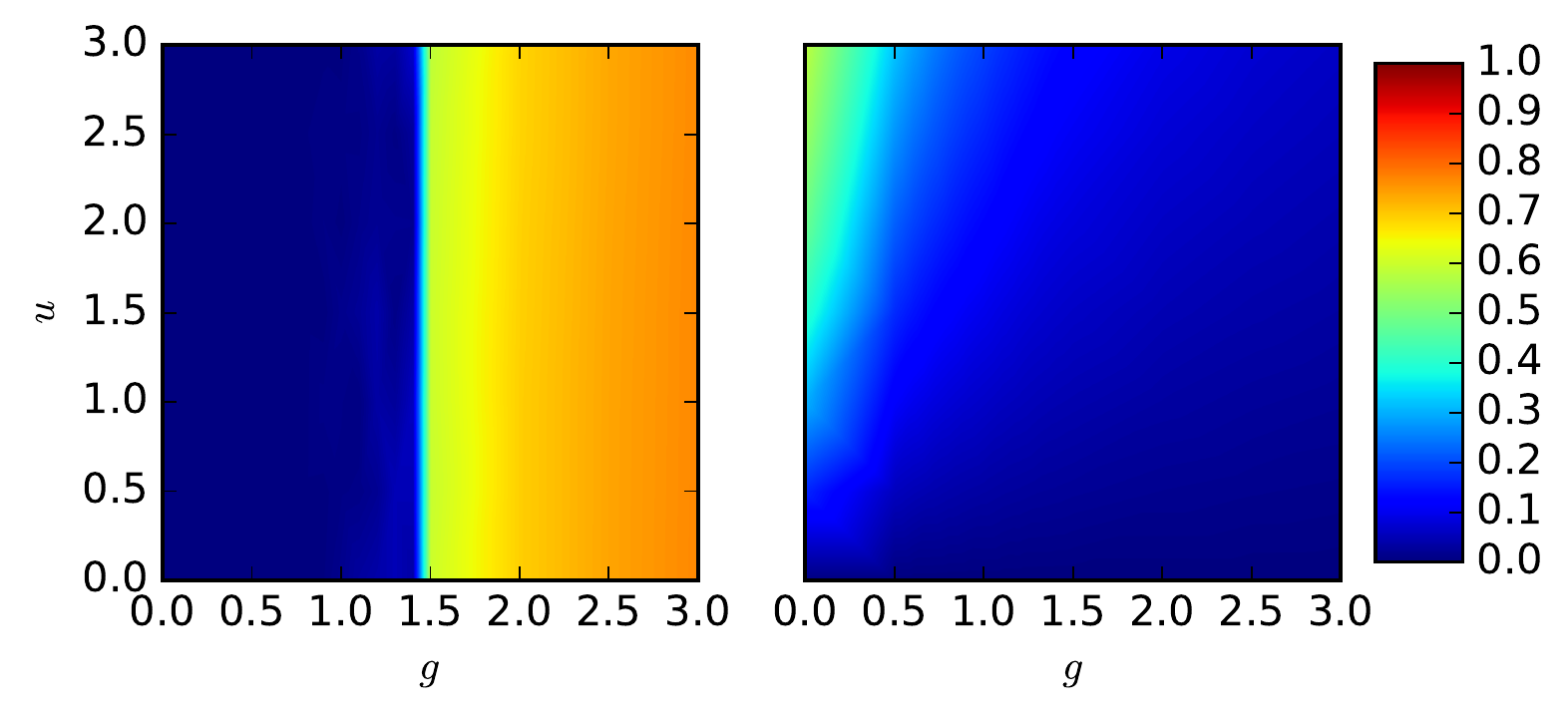}
\caption{The in-plane polarization $\phi_\text{pol}$ (left panel) and transverse polarization $\phi_z$ (right panel) vs. $g$ and $u$ for a system of 48 dipoles on a triangular lattice with $\beta = 5.1$ $(2\pi B)^{-1}$ and $\tau = 0.0375$ $(2\pi B)^{-1}$.}
\label{fig:tri_z}
\includegraphics[width=4in]{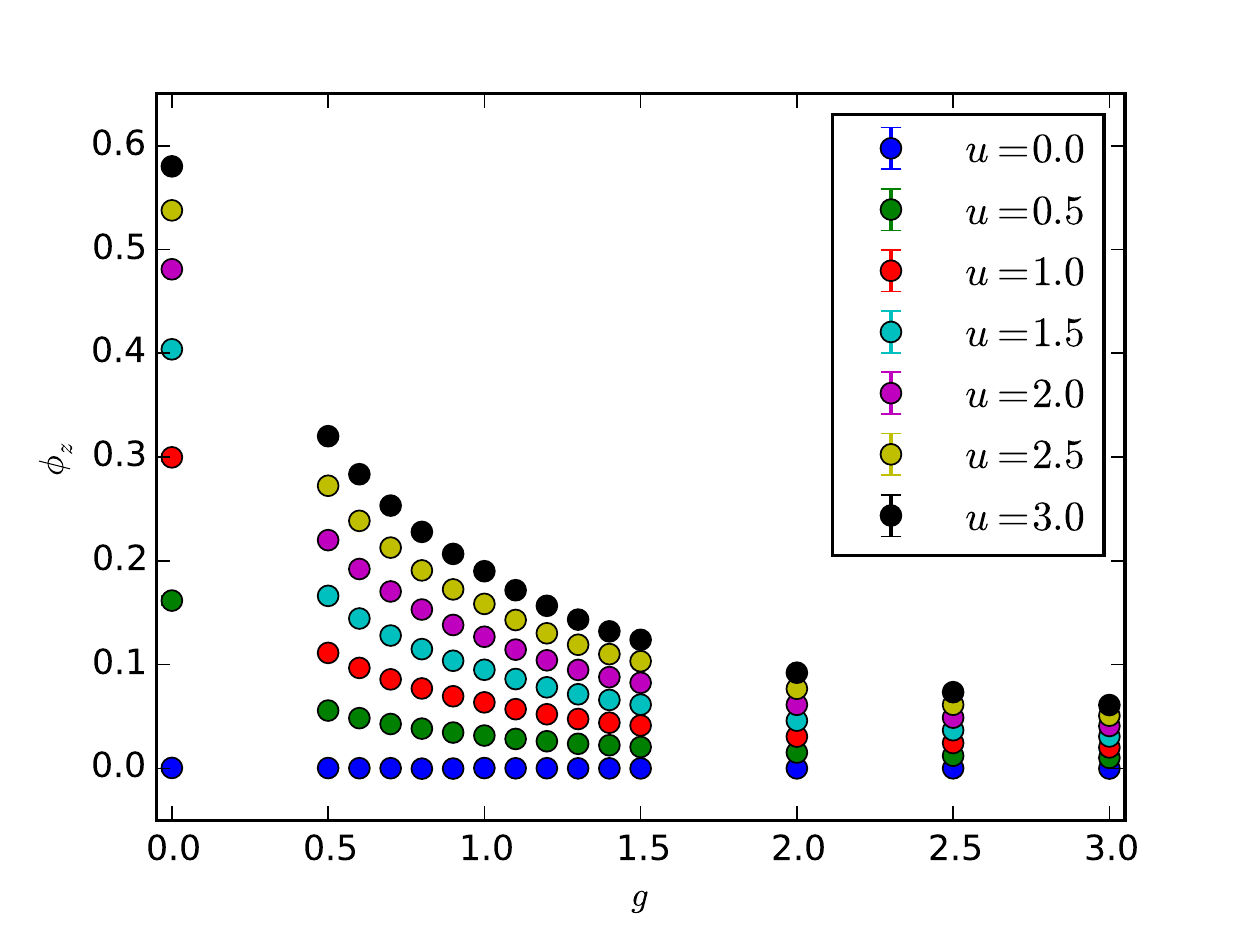}
\caption{Transverse polarization, $\phi_z$, vs. $g$ for a system of 48 dipoles on a triangular lattice with $\beta = 5.1$ $(2\pi B)^{-1}$ and $\tau = 0.0375$ $(2\pi B)^{-1}$. This shows the same data as the right panel in Figure~\ref{fig:tri_z} but presented in a way to highlight the gradual decay of the $z$-polarization as the interaction strength $g$ is increased.}
\label{fig:tri_z_slice}
\end{figure}

The right and left panel of Figure~\ref{fig:tri_z} show the dependence of the transverse and in-plane polarizations, $\phi_z$ and $\phi_{pol}$, respectively, on the parameters $g$ and $u$.  We see that there is a decrease in the $z$ polarization (right panel), both as $g$ is increased and as $u$ is decreased. The fact that the $z$ polarization decreases as $u$ is decreased is not surprising, since a decline in the polarization with decreasing applied field is generally expected. The dependence on interaction strength $g$ is less immediately intuitive.
However for classical dipoles, one can show that in the limit $g \gg 1$, depending on the lattice geometry, the minimal classical energy configuration can have all dipoles polarized in the plane of the lattice with zero polarization out of the plane. The decline in the out of plane polarization for the PIGS results as $u$ decreases appears to be gradual with respect to both $g$ and $u$, with no sharp drop-off. This gradual decrease can be seen more clearly in Figure~\ref{fig:tri_z_slice}.

The behavior of the in-plane $\phi_\mathrm{pol}$ evident in the left panel of Figure~\ref{fig:tri_z}, is quite different, showing a very sharp rise in $\phi_\mathrm{pol}$ at a critical value $g \approx 1.5$ that is very nearly independent of $u$. This transition occurs over a very narrow range of $g$ values, and does not coincide with the decrease in $\phi_z$. While there is a slight modulation of the position of this crossover behavior with respect to $u$, the precise location of the transition is nevertheless difficult to pin down in this range of $u$ values, with the variation with $u$ being less than the width of the crossover region for the lattice sizes considered here.

This sharp variation of the in plane polarization with respect to the interaction strength is highly suggestive of a quantum phase transition between an essentially unpolarized state and a state exhibiting macroscopic polarization. In light of the behavior of one dimensional lattice systems of rotors~\cite{Abolins2011} this is not altogether surprising. In fact, in one dimensional systems without electric fields this crossover from an unpolarized state to a polarized state occurs at a similar value of $g$. The large statistical fluctuations near the critical value of $g$ precludes a definite answer whether $\phi_\mathrm{pol}$ varies continuously
with $g$ (second order phase transition) or has a jump (first order phase transition).  But a continuous
second order phase transition would be consistent with the large statistical fluctuations, and
in particular with the slow convergence
of $\phi_\mathrm{pol}$ close to the critical value of $g$ when the imaginary time path length
$\beta$ is increased.  This is discussed in detail in the Appendix.
What is somewhat surprising is that this behavior in two dimensions on a triangular lattice appears to be only weakly dependent on the strength $u$ of the external field, at least for moderately strong external fields. This suggests that this behavior can have profound effects on systems over a wide range of experimentally accessible external field strengths.

\subsection{PIGS Results for dipoles on Square Lattices}
\label{subsec:sq_results}

\begin{figure}[h]
\centering
\centering
\includegraphics{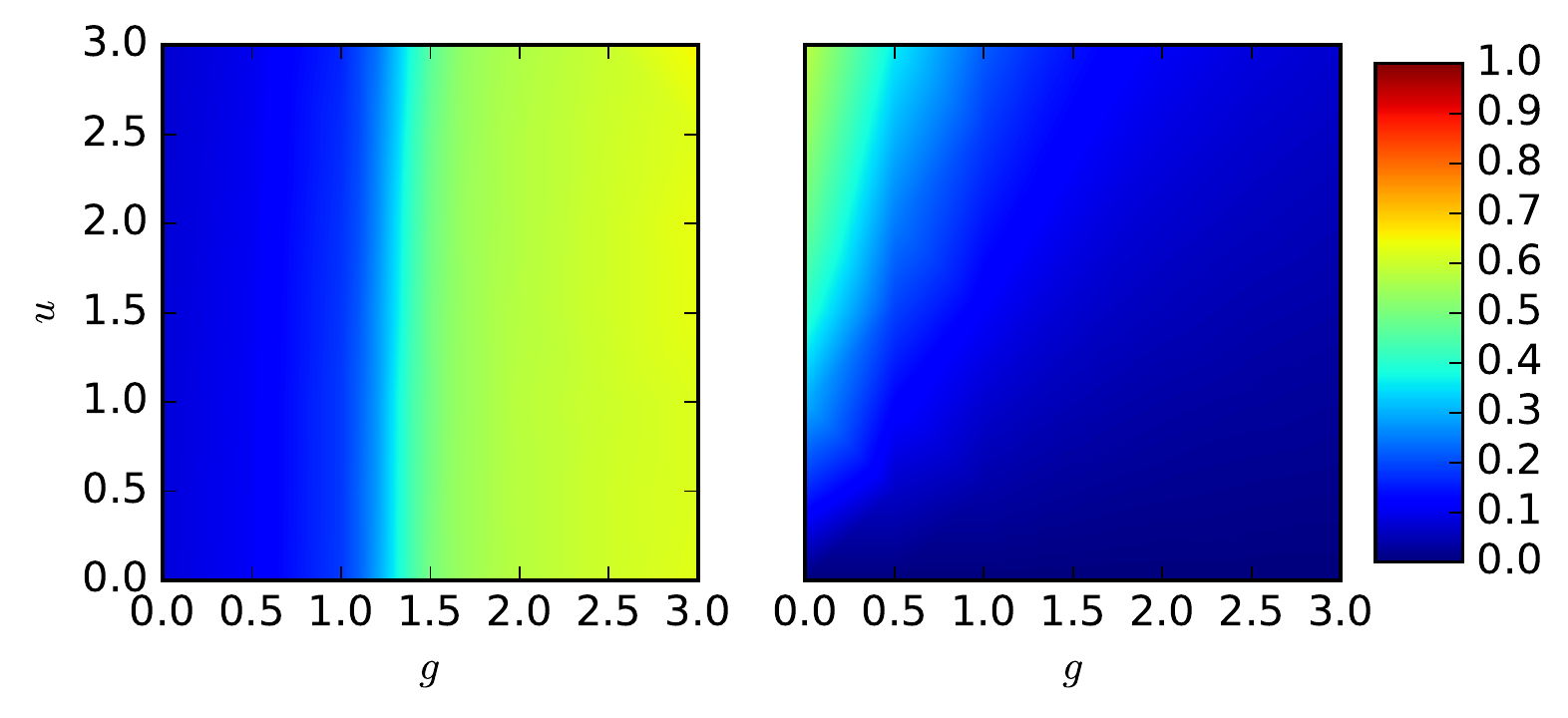}
\caption{The striped order parameter $\phi_{xy}$ (left panel) and transverse polarization $\phi_z$ (right panel) vs. $g$ and $u$ for a system of 64 dipoles on a square lattice with $\beta = 4.2$ $(2\pi B)^{-1}$ and $\tau = 0.0375$ $(2\pi B)^{-1}$.}
\label{fig:sq_z}
\end{figure}

Systems of 64 dipoles confined to square lattices were simulated using the PIGS method, varying both $u$ and $g$ 
over the range 0 to 3. The time step was $\tau = 0.0375$ $(2\pi B)^{-1}$ and the imaginary time path length
was $\beta = 4.2$ $(2\pi B)^{-1}$. For a discussion of the convergence with $\tau$ and $\beta$ we refer again to the Appendix.
The dependence of the transverse polarization $\phi_z$ in the case of a square lattice is shown in the right panel of Figure~\ref{fig:sq_z}, showing very similar behavior to that shown for the transverse polarization on a triangular lattice in the right panel of Figure~\ref{fig:tri_z}. The reasoning behind the trends in $\phi_z$ with $g$ and $u$ 
is identical to that described for the triangular lattice. As with the triangular lattice results, the decline in $\phi_z$, shown in the right panel of Figure~\ref{fig:sq_z}, is gradual and not sharp.

The left panel of Figure~\ref{fig:sq_z} shows the $g$ and $u$ dependence of the in-plane striped order parameter, $\phi_{xy}$ for the square lattice. Though the ordering described in this plot is quite different to the in-plane ordering shown in the left panel of Figure~\ref{fig:tri_z} for the triangular lattice, {\em i.e.} showing striped ordering rather than a totally polarized ordering, the qualitative behavior exhibited is very similar,
with a sharp transition from a disordered phase to an ordered phase at a critical value of $g$ around $1.25$. Similar to the triangular lattice, there is a slight trend toward a higher critical value of $g$ as the strength $u$ of the transverse field is increased. It should be noted that the critical value of $g$ for the square lattice differs from the corresponding
value for the triangular lattice, although they are quite similar.

\subsection{Mean Field Analysis}
\label{subsec:meanfield}

To compare with the PIGS results, we also studied the phase diagrams using a self-consistent mean field theory. In this 
approach the $N$-body wave function 
is approximated by a product of single-body wave functions
\begin{equation}
\ket{\Psi_\text{mf}} = \prod_{i=1}^{N} \ket{\phi_i}. \label{eq:mf_wf}
\end{equation}
Under this approximation the expectation energy of Eq.(\ref{eq:ham_simp}) can be written as
\begin{align}
\expect{E} &= \bra{\Psi_\text{mf}} H \ket{\Psi_\text{mf}} \\
	&= \sum_{i=1}^{N} \bra{\phi_i}\mathbf{L}_i^2 - u \mathbf{n}_i \cdot \hat{\mathbf{e}}\ket{\phi_i} + \frac{g}{2} \sum_{i = 1}^N \sum_{j = 1}^N \bra{\phi_i} \mathbf{n}_i \ket{\phi_i} \cdot \mathbf{S}_{ij} \cdot \bra{\phi_j} \mathbf{n}_j \ket{\phi_j} \\
	&= \sum_{i=1}^N \bra{\phi_i} h_\text{eff}(i) \ket{\phi_i}.
\end{align}
Expanding the single particle wave functions in the basis of spherical harmonics centered on particle $i$,
\begin{equation}
\ket{\phi_i} = \sum_{l=0}^\infty\sum_{m=-l}^l c_{i,lm} \ket{lm},
\end{equation}
and minimizing the expectation value of the energy with respect to the expansion coefficients, $c_{i,lm}$, 
while requiring self-consistency, provides an approximation to the ground state energy and associated wave function. A value of $l_{max}=4$ was found sufficient 
to converge the energies and order parameters over the range of $g$ and $u$ studied here. 

\begin{figure}[h]
\centering
\includegraphics{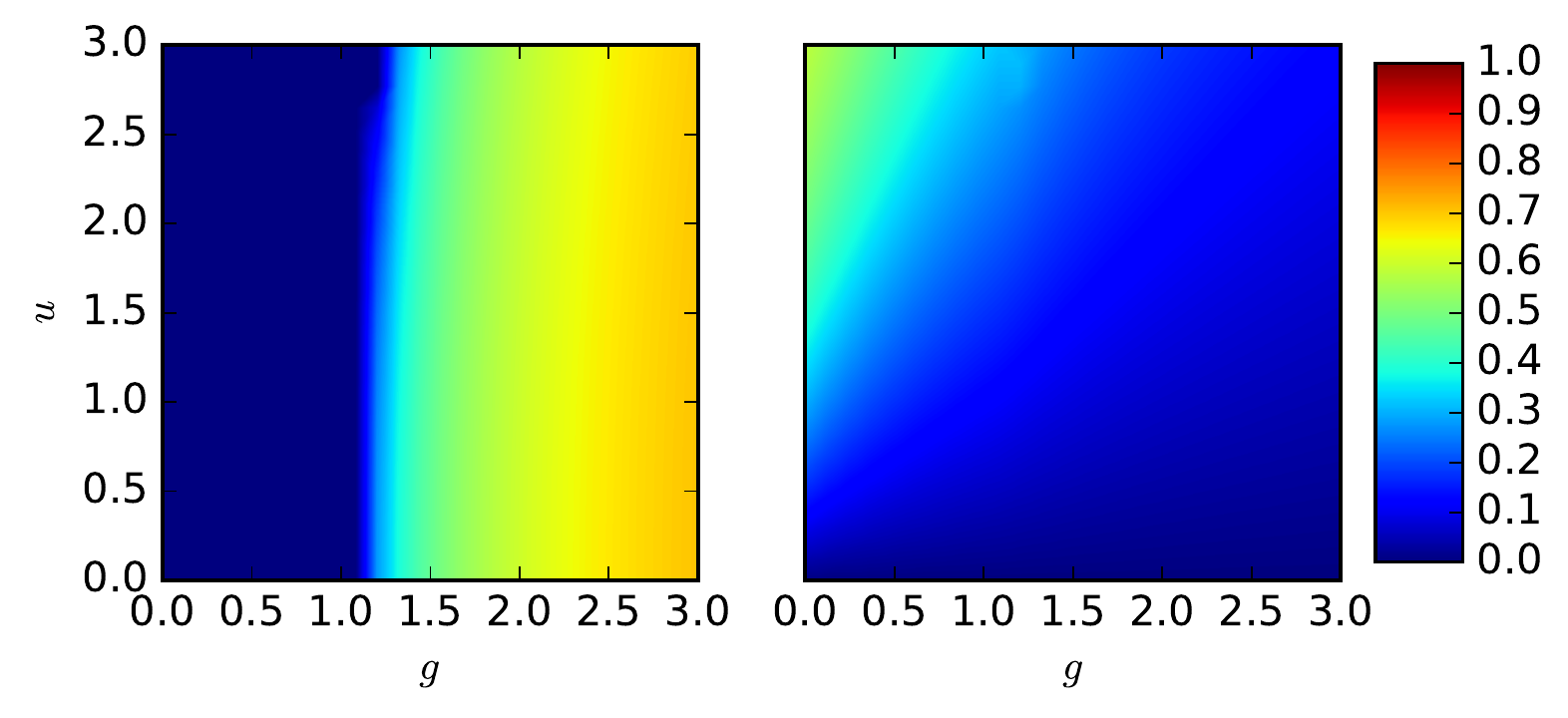}
\caption{Mean field calculation of the in-plane order parameter $\phi_\text{pol}$ (left panel) and transverse polarization $\phi_z$ (right panel) as a function of $g$ and $u$ for 12 dipoles on a triangular lattice.}
\label{fig:tri_xy_mf}
\end{figure}

The left and right panels of Figure~\ref{fig:tri_xy_mf} show, respectively, plots of the order parameters
$\phi_\mathrm{pol}$ and $\phi_z$
as functions of the field strength $u$ and interaction strength $g$, for 12 dipoles on a triangular lattice, with a spatial cutoff for the range of the periodic sum of 100 nearest-neighbor distances and a maximum angular momentum $l_\text{max} = 4$, corresponding to 300 total basis functions, derived from the self-consistent field wave function, $\ket{\Psi_\text{mf}}$. 

\begin{figure}[p!]
\centering
\includegraphics{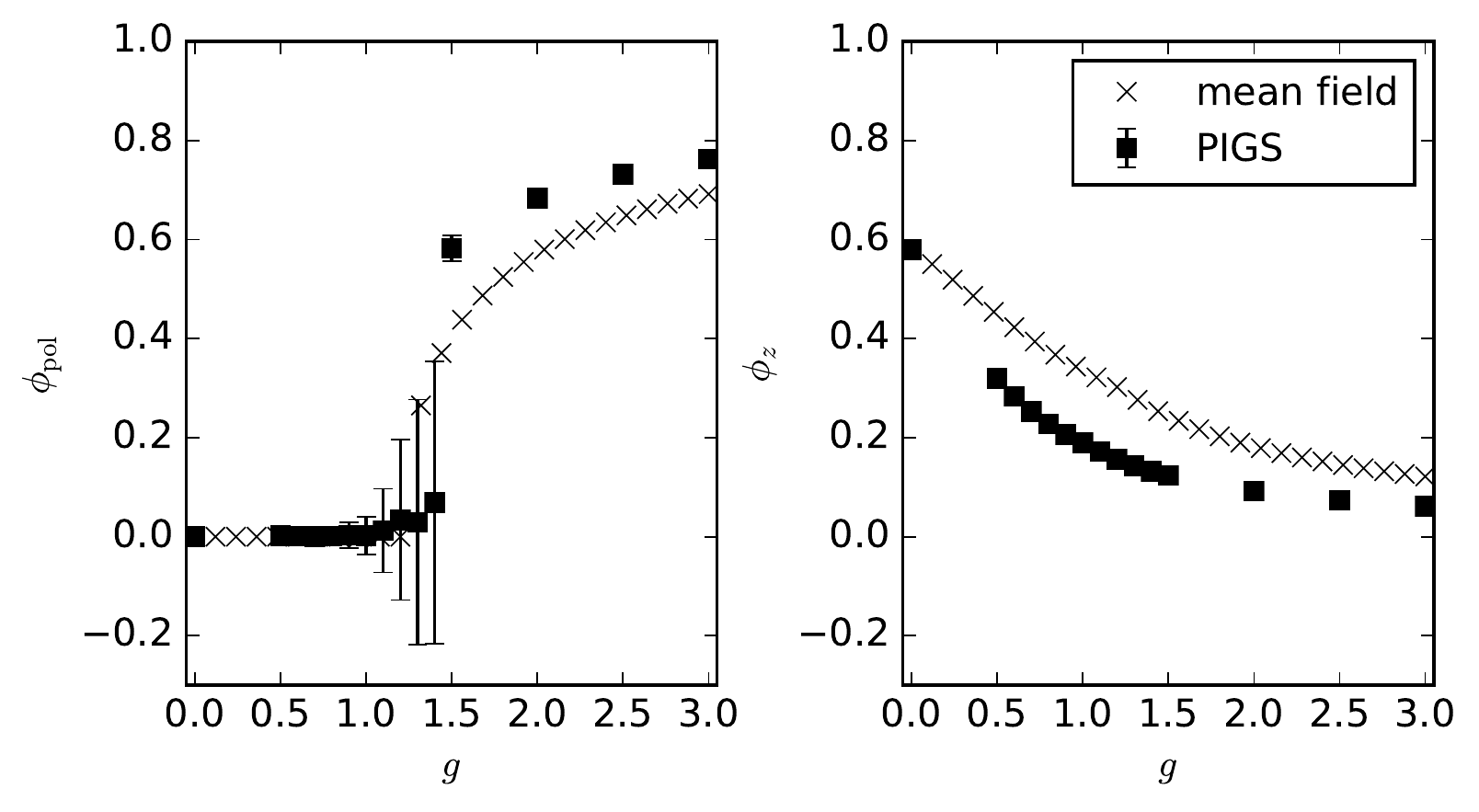}
\caption{Comparison of $\phi_\mathrm{pol}$ (left  panel) and  $\phi_z$ (right panel) on triangular lattices calculated from the mean field approximation (12 dipoles) and from PIGS (48 dipoles) as a function of $g$ at $u = 3$.}
\label{fig:tri_pol_comp}
\includegraphics{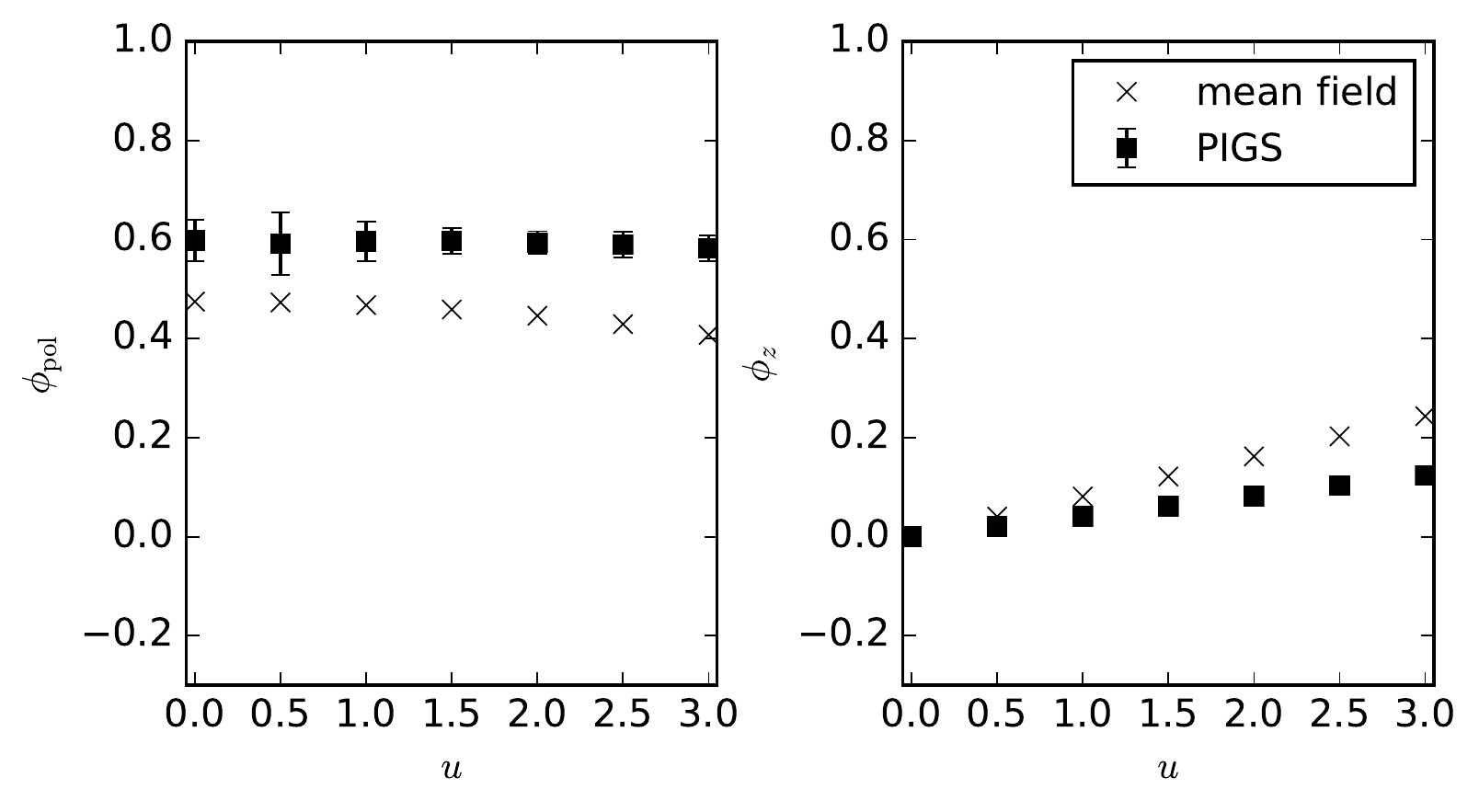}
\caption{Comparison of $\phi_\mathrm{pol}$ (left panel) and $\phi_z$ (right panel) on triangular lattices calculated from the mean field approximation (12 dipoles) and from PIGS (48 dipoles) as a function of $u$ at $g = 1.5$.}
\label{fig:tri_z_comp}
\end{figure}

While the mean-field results show the same overall features as the Monte Carlo results, they show 
quantitative differences arising from the neglect of correlation between particles in the mean field approach.
Only in the case of non-interacting dipoles ($g=0$), do the mean field results become exact.
The differences between the mean-field and PIGS results are quantified in Figures~\ref{fig:tri_pol_comp} and ~\ref{fig:tri_z_comp}, which show cuts of $\phi_\mathrm{pol}$ and $\phi_z$ at fixed transverse field strength $u$ and fixed interaction strength $g$, respectively.  
Each of these cuts show significant differences between the mean field and PIGS values, with the mean field values lying systematically below the PIGS values over a range of values of $u$ and $g$ for the in-plane polarization $\phi_\mathrm{pol}$, and systematically above the PIGS values for the transverse polarization $\phi_z$.
We can qualitatively understand the different sign of the deviation of
the mean field results for $\phi_\mathrm{pol}$ and $\phi_z$ from those obtained by PIGS in terms of the dipole correlations.
A nonzero value of $\phi_\mathrm{pol}$ can result only from the interactions
between dipoles, which cause correlations between these.  Since such correlations are
neglected in the mean field approximation, this underestimates $\phi_\mathrm{pol}$. In contrast,
$\phi_z$ results from the external field, which does not cause correlations.  $\phi_z$ must
exhibit the opposite trend of of $\phi_\mathrm{pol}$: if there is less in-plane order as
quantified by $\phi_\mathrm{pol}$, this increases the out-of-plane components
of the orientation vectors of the dipolar rotor, thus allowing an increase of $\phi_z$.
The underestimation of $\phi_\mathrm{pol}$ in the mean field approximation
is therefore accompanied by an overestimation of $\phi_z$.

A mean-field analysis analogous to that undertaken for triangular lattices was also performed for square lattices. The differences between the mean field and the  PIGS simulation results for the square lattice are qualitatively similar to that observed with triangular lattices, showing the same general trends as those in Figures~\ref{fig:tri_pol_comp} and \ref{fig:tri_z_comp} and can be rationalized by similar arguments as above.

\section{Discussion}
\label{sec:discussion}

Because of the 
anisotropic nature of the dipole-dipole interaction, the predicted 
classical orientational orderings of 
ground states of dipolar ensembles on square and triangular lattices in two dimensions is quite different. 
The quantum Monte Carlo results presented here show that dipoles on a triangular lattice prefer orderings featuring a net in-plane polarization of the system, with all such polarized phases being degenerate in the absence of any fields in the plane of the lattice. On a square lattice, dipoles are predicted to adopt a striped ordering, characterized by no 
average net polarization in the lattice plane. These differences can be explained by differences in the symmetry of the lattices: the layout of the triangular lattice results in 6 nearest neighbor interactions for each dipole and 12 next-nearest neighbor interactions, while dipoles on the square lattice possess only 4 nearest and 4 next-nearest neighbors.
Because of differences in the lattice geometries, the number of neighbors in successive shells, as well as the distance between successive shells, differs between lattices, leading to different contributions to the
long-ranged dipole-dipole interaction.  Its anisotropy causes to partial cancellation of
positive and negative contributions, which leads to the different phases favored by the dipole interaction for
the two lattices.

\begin{figure}[h]
\centering
\includegraphics{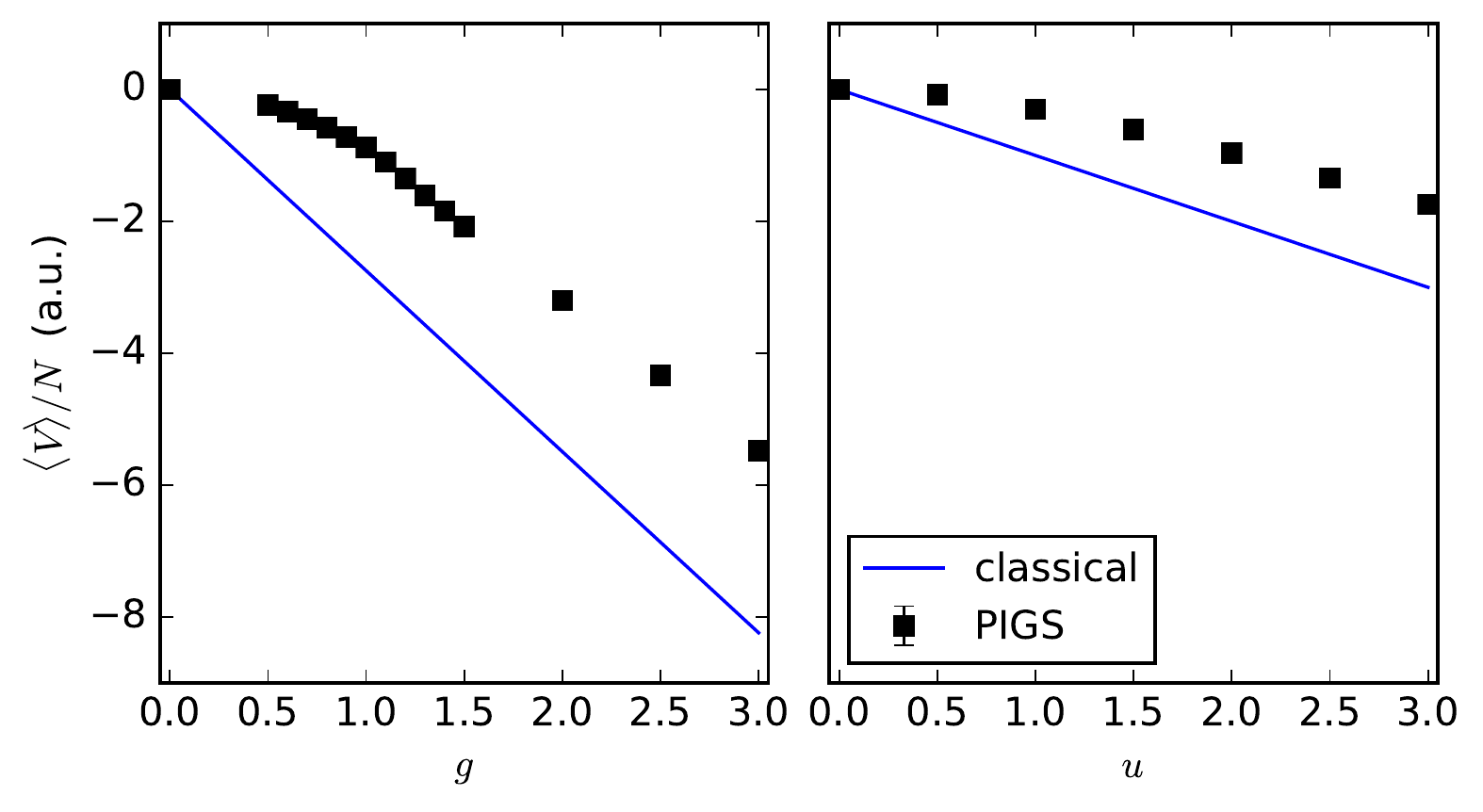}
\caption{Comparison of the potential energy per particle from PIGS simulations of 48 dipoles on a triangular lattice, with the classical potential energy of a system of 675 dipoles on the same lattice, polarized in both cases along the $x$-axis. The potential energy is shown as a function of the dipole-dipole interaction strength parameter $g$, at $u = 0$ (left panel) and as a function of the strength of the electric field $u$, at $g=0$ (right panel).}
\label{fig:quantum_PEcomparison}
\end{figure}

We stress that dipoles situated on a regular lattice described by the
Hamiltonian (\ref{eq:ham}) are a proper quantum many-body system, and the
orientational ordering is a quantum phase transition.  The classical
limit at zero temperature (no rotational kinetic energy) is taken
by letting $B \to 0$, which in our energy units corresponds to
$g\to\infty$ and $u\to\infty$.  In the classical limit the dipoles
orient themselves such that the total potential energy, consisting of the
interaction and the external field, is minimized.  At $T=0$ K, classical dipoles
are therefore in an ordered phase for any nonzero value of $g$ and $u$.
The correct quantum description accounts for the quantum kinetic energy
which leads to a higher potential energy due to orientational delocalization.
As a measure of ``quantumness'',
we use the difference between the potential minimum $V_{\rm min}$ and the quantum
mechanical expectation value of the potential $\langle V\rangle$ calculated in our
PIGS simulations.  Figure\ref{fig:quantum_PEcomparison} shows $V_{\rm min}$ per particle
(line) and the PIGS expectation value $\langle V\rangle$ (symbols) per particle
for the triangular lattice.  
In the left panel, the strength of the external potential $u$ is varied,
with $g=0$, and in the right panel the strength of the interaction $g$
is varied with $u=0$.  $V_{\rm min}$ is a straight line
because we simply scale the respective potential and thus the minimum
of the potential.  The ratio between $\langle V\rangle$ and
$V_{\rm min}$ can be regarded as measure of quantumness.  For small
potential strengths $g$ or $u$, the quantum kinetic energy (not shown)
is dominant leading to a large delocatization and thus an expectation
value $\langle V\rangle$ much higher and close to zero.  As $g$ or $u$
are increased, the potential becomes more dominant.  For $g\to\infty$
or $u\to\infty$ the ratio $\langle V\rangle / V_{\rm min}$ will converge
to unity, as expected, and in this limit the quantum system is in an ordered
phase like the classical system.  However, we note in the whole range
of $g$ and $u$ studied in this work, the ratio is significantly less
than unity.  Particularly, in the vicinity of the phase transition
with a critical value $g\approx 1.5$ (see the left panel of Figure~\ref{fig:tri_z}), the
left panel of Figure~\ref{fig:quantum_PEcomparison}
shows that $\langle V\rangle / V_{\rm min}\approx 0.5$.  In the interesting
regime around the quantum phase transition to an ordered phase,
dipolar rotors on a lattice require a full quantum mechanical description.

Our PIGS results reveal that, including the full quantum effects, dipoles on a triangular lattice will still tend to form a polarized phase at sufficiently high interaction strength
$g$. Furthermore this tendency appears to be nearly independent of the applied field strength in the transverse direction.
Since some proposals of self-assembled two-dimensional crystals rely on 
imposition of transverse fields to ensure dipole orientation~\cite{Buchler2007,Astrakharchik2007}, such a phase transition could limit the densities at which these two-dimensional crystals are stable. This is because at higher densities, which is to say higher values of $g$, dipoles will naturally tend to form attractive head-to-tail configurations, rather than the repulsive transverse polarized configurations 
that lead to stable crystals, in which case the dipoles may cease to be trapped due to collisions~\cite{Ni2008}.

\begin{figure}[h!]
\centering
\includegraphics{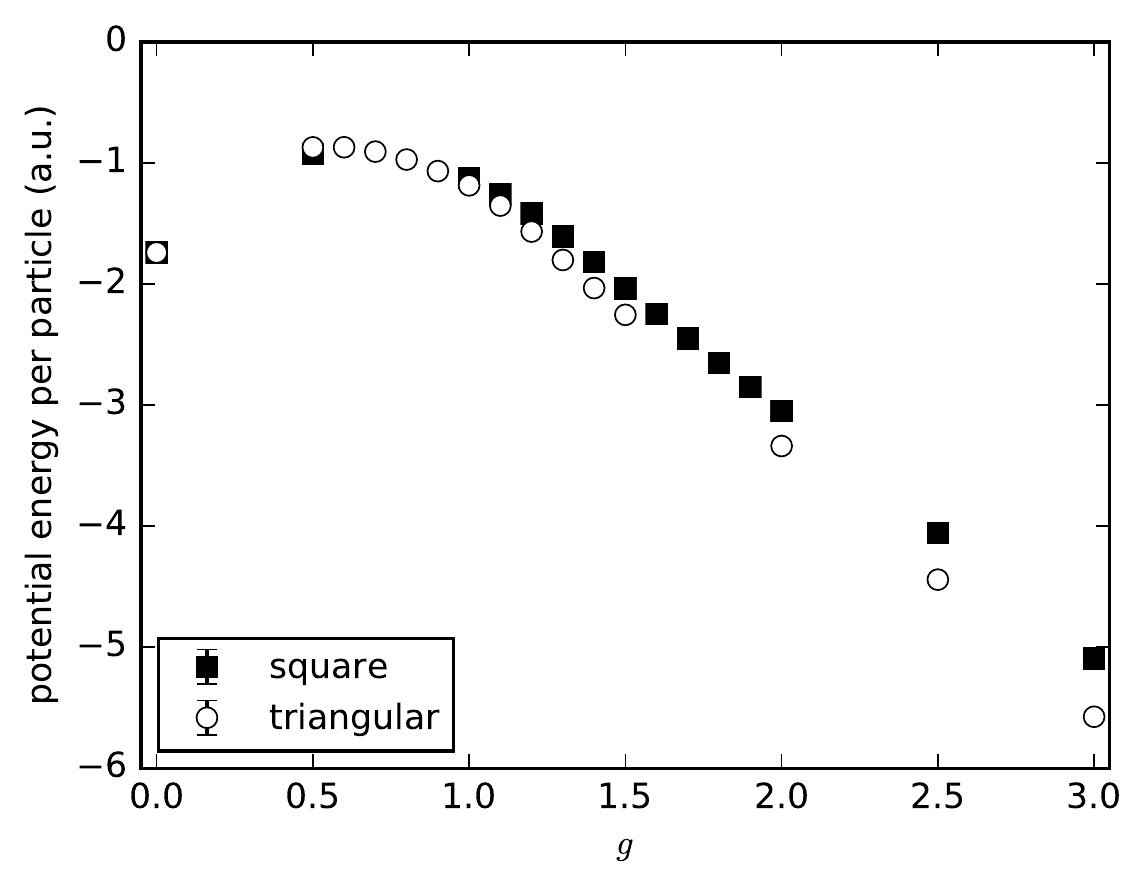}
\caption{Comparison of the potential energy per particle from PIGS simulations of 64 dipoles on a square lattice and 48 dipoles on a triangular lattice at $u = 3$. At higher values of the interaction strength, $g$, the potential energy per particle on a triangular lattice is found to be marginally lower than in the equivalent case on a square lattice, possibly helping to explain differences in the value of $g$ at which the transition from a disordered state to an ordered state occurs.}
\label{fig:pot_en_comp}
\end{figure}

In contrast, when confined to square lattices, dipoles will tend to form striped phases at high values of $g$. Despite differences in the precise nature of the orientational ordering at the higher values of $g$, the location in $g$ at which the transition occurs is remarkably similar on the two lattices. The differences in the orientational orderings could potentially derive from a variety of reasons. One possible reason is differences in system size, since both systems exhibited slight finite size effects in the PIGS calculations. Differences in the potential energy per particle between the fully polarized state (on the triangular lattice) and the striped phase (on the square lattice) could also lead to differences in the precise value of the interaction strength, $g$, at which the transition occurs on each lattice. For a given $g$ value, the potential energy per particle of the fully polarized state on the triangular lattice is predicted to be marginally lower than that of the striped phase on the square lattice, a trend which carries over to the PIGS results and can be seen in Figure~\ref{fig:pot_en_comp}. This suggests that the value of $g$ required to fully polarize a system of dipoles on a triangular lattice should be smaller than that required to form a striped phase on a square lattice, which is precisely what was observed in the Monte Carlo calculations.
 
\section{Conclusions}
\label{sec:conclusions}

Since the interaction strength $g$ includes the effect of not only the dipole moment, $d$, but also of the average inter-particle distance, the results found in this work imply that such ordered phases should be expected to occur for sufficiently dense systems. 
From the values in Table~\ref{Table1:dipole parameters}, it can be seen that for diatomics of potential interest to trapped cold molecule experiments, having permanent dipole moments between 5 and 10 Debye, the transition from weakly  to strongly interacting behavior with onset of marked in-plane polarization is expected to occur for inter-particle distances on the order of 10 - 30 nm. The best candidates for observing these phases are molecules that also have small rotational constants, i.e., the heavier species.  In particular, CsI achieves $g = 1$ at a lattice spacing as large as approximately 30 nm.
Experimental systems that have been realized to date have been located well within the realm of 
the weakly interacting regime of unpolarized dipoles, having been formed at much lower densities~\cite{Ni2008,Yan2013}. Additionally, even predicted self-assembled crystalline phases of dipolar molecules are expected to occur in the low density, {\em i.e.} the low interaction strength, regime~\cite{Buchler2007}. This should allow these self-assembled crystals to be probed without fear of entering a phase with in-plane ordering which in the absence of an external lattice potential, would lead to a collapse of the observed crystal due to attractive interactions.

This effect may also be important in systems of dipolar coupled pseudo-spins or excitons within a strongly interacting regime~\cite{Kocherzhenko2014}. Such a strongly interacting regime of dipolar coupled pseudo-spins~\cite{Kocherzhenko2014} is precisely the same as the high density regime where the dipoles are likely to
interact strongly with one another.
More directly relevant to dipolar molecules, as was pointed out in Section~\ref{sec:intro}, new trapping schemes~\cite{Ritt2006,romeroisartPRL13,Nascimbene2015,Lkacki2016,Perczel2017} may lead to lattices with much smaller lattice constants than are currently implemented by optical lattices.
In the present study the limits of stability in the strong interaction regime were not truly probed, as the dipolar particles were treated as translationally frozen with no explicit spatial trapping potential. 
Further study is warranted in this regard, since the dynamics of an unstable ensemble trapped by finite optical lattice potentials would require overcoming or tunneling through the additional optical lattice potential. However, the results of this study show that at the very least it would seem that molecules can not be reliably counted upon to remain transversely polarized in the high density limit, even in the presence of strong external potentials in the transverse direction. 

\section{Acknowledgements}

B. P. A. and K. B. W. were supported by funding from the National Science Foundation Grant No. CHE-1213141. R. E. Z. acknowledges funding from the Austrian science fund FWF (grant number P23535-N20).

\begin{appendix}

\section{Parameter Convergence}
\label{subsec:parameter_convergence}

The PIGS method is an exact quantum Monte Carlo method for bosons, apart from the bias of finite time step, $\tau$, and
finite imaginary time path lengths, $\beta$.  The bias can be made arbitrarily small by choosing small values for
$\tau$ and large values for $\beta$.
To benefit from the systematic nature of the approximations it is necessary to carry out simulations at varying $\tau$ and $\beta$. In general such studies must be carried out at different points in the space of parameters in the Hamiltonian of interest, in the present work this is the values of the parameters controlling the interaction strength with an external field, $u$ in Eq.~(\ref{eq:ham}), and the strength of the dipole-dipole interaction, $g$ in Eq.~(\ref{eq:ham}).

\begin{figure}[p!]
\centering
\includegraphics{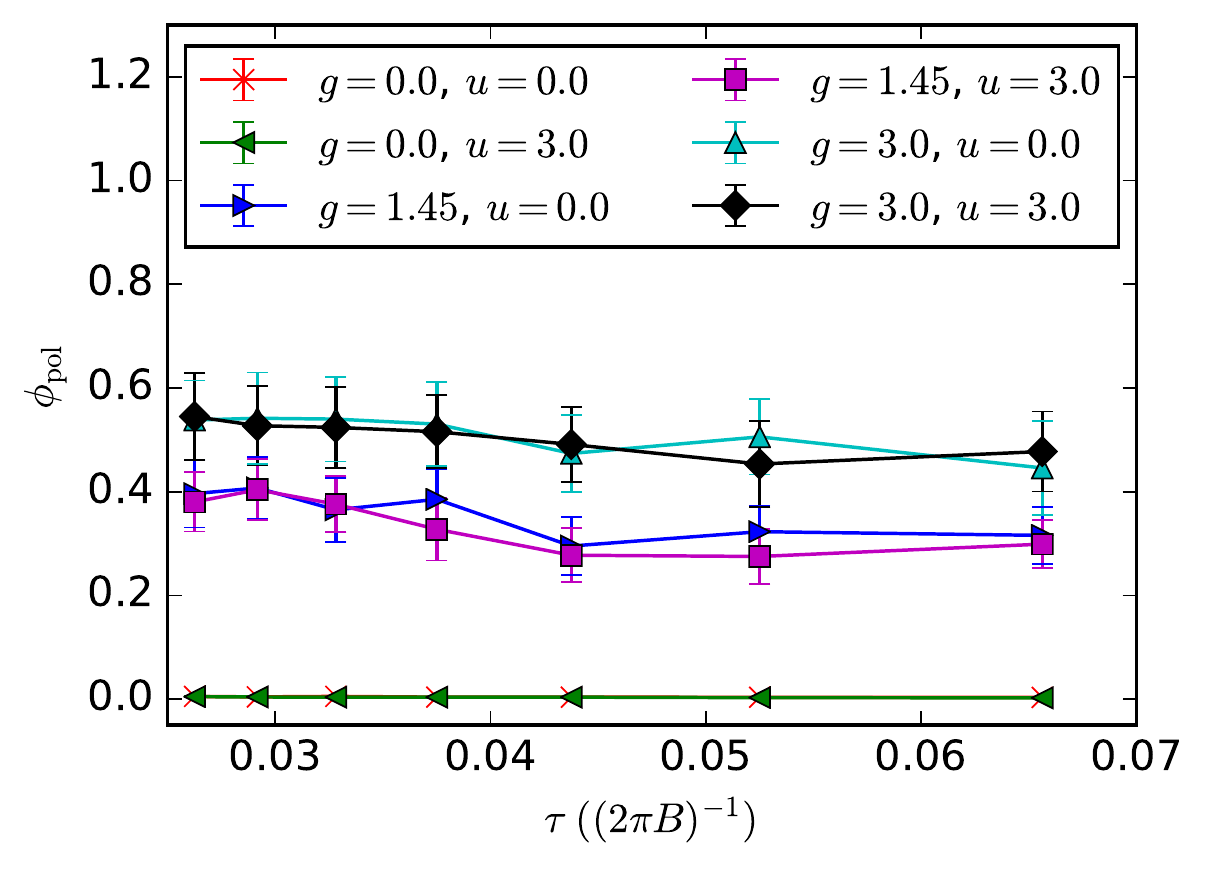}
\caption{The behavior of $\phi_\mathrm{pol}$ as a function of the small time step, $\tau$, for a system of 48 dipoles on a triangular lattice at a constant imaginary path length, $\beta = 4.2$ $(2\pi B)^{-1}$.}
\label{fig:triangular_tau}
\includegraphics{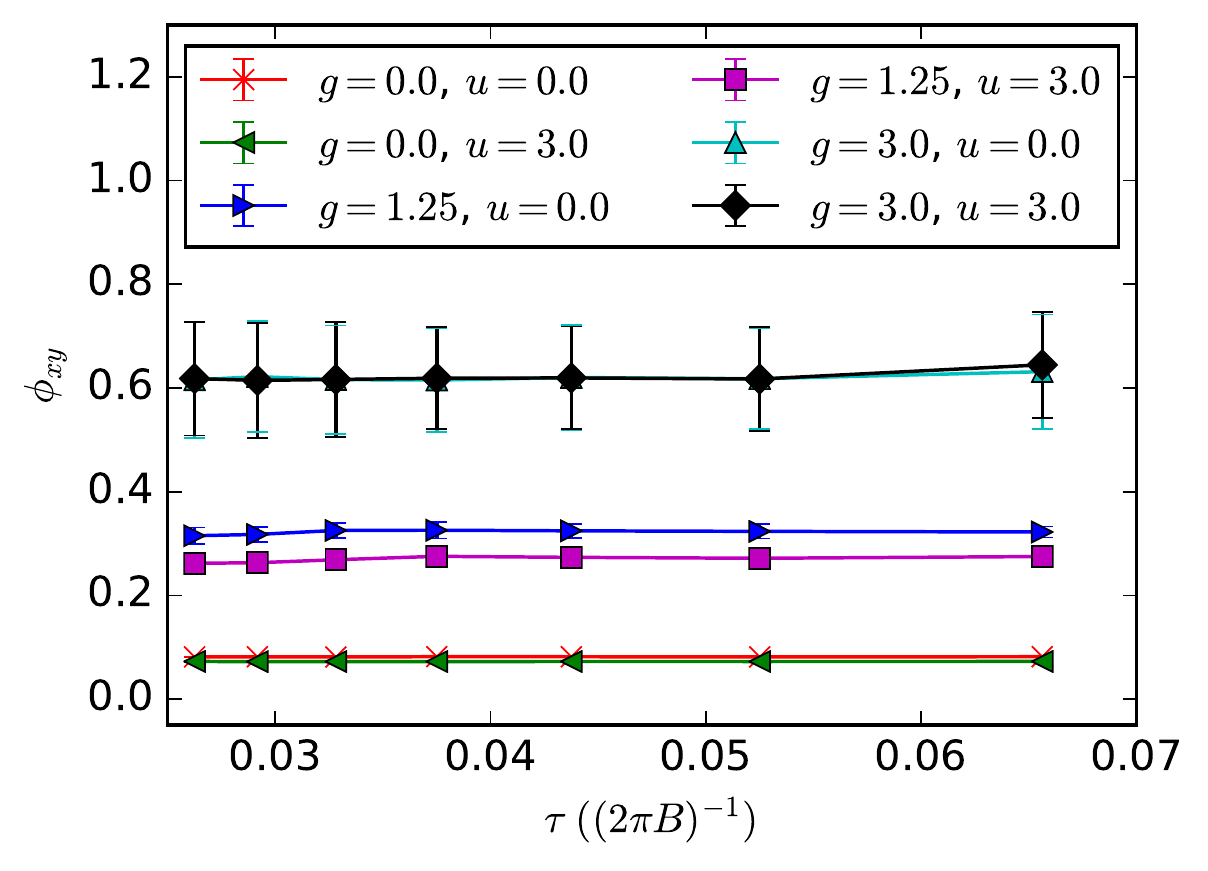}
\caption{The behavior of $\phi_{xy}$ as a function of the small time step, $\tau$, for a system of 64 dipoles on a square lattice at a constant imaginary path length, $\beta = 4.2$ $(2\pi B)^{-1}$.}
\label{fig:square_tau}
\end{figure}

\begin{figure}[p!]
\centering
\includegraphics{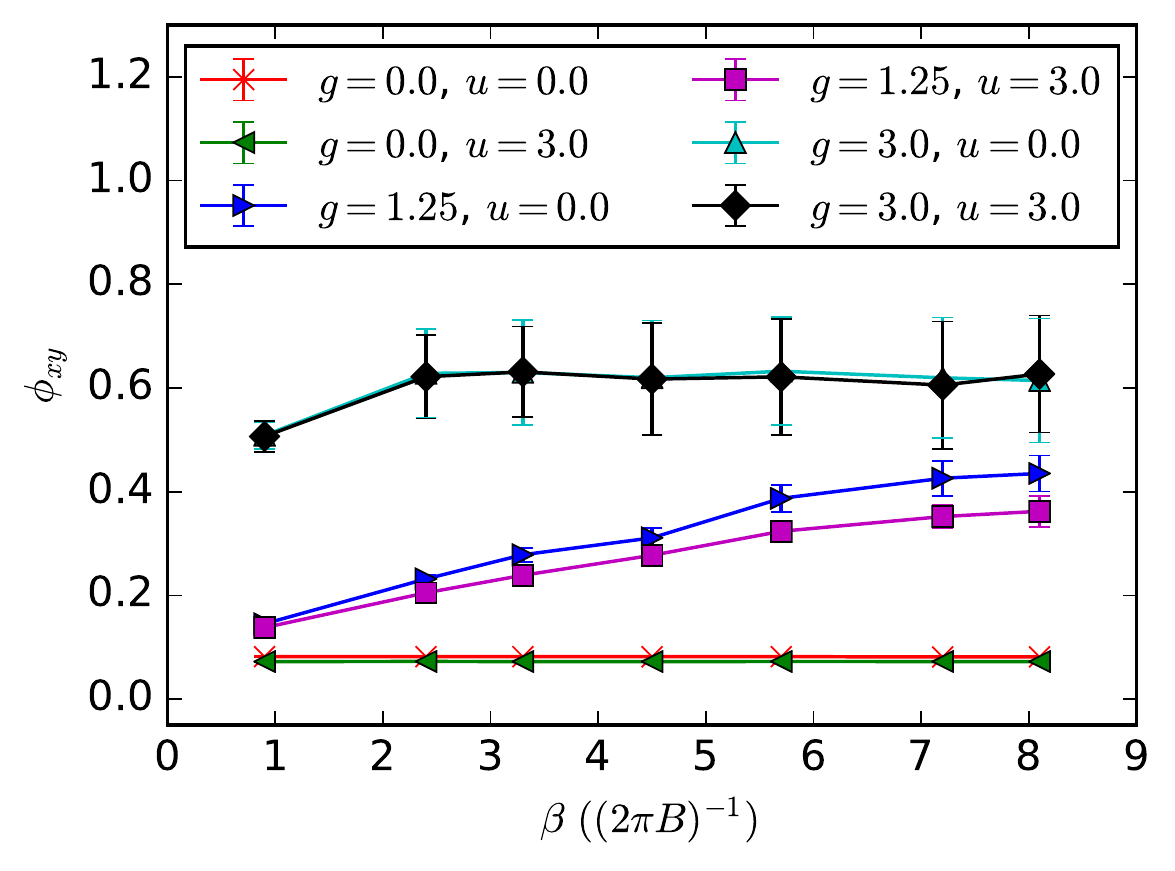}
\caption{The behavior of $\phi_{xy}$ as a function of the imaginary path length, $\beta$, for a system of 64 dipoles on a square lattice at a constant small time step, $\tau = 0.0375$ $(2\pi B)^{-1}$.}
\label{fig:square_beta}
\includegraphics{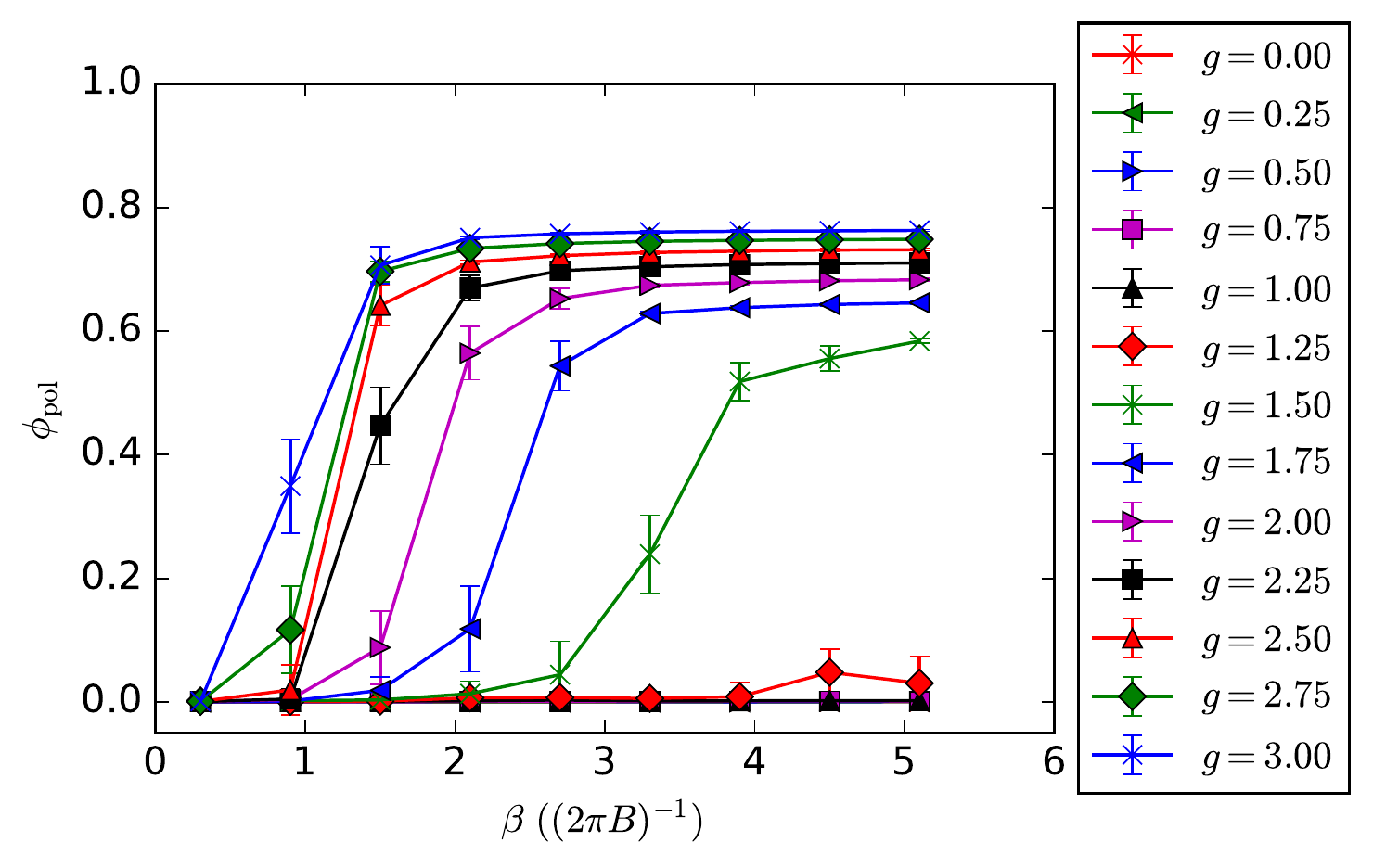}
\caption{The behavior of $\phi_\mathrm{pol}$ as a function of the imaginary path length, $\beta$, for a system of 48 dipoles on a triangular lattice at a constant small time step, $\tau = 0.0375$ $(2\pi B)^{-1}$ and constant $u = 0.5$.}
\label{fig:triangular_beta}
\end{figure}

Figure~\ref{fig:triangular_tau} shows the behavior of the order parameter $\phi_\mathrm{pol}$ as a function of the time step, $\tau$, at constant path length, $\beta = 4.2$ $(2\pi B)^{-1}$ for six different points in the space of Hamiltonian parameters for dipoles on a triangular lattice with 48 dipoles in the periodic simulation cell. These points correspond to the limits of the present study with two additional points at intermediate dipolar interaction strength, $(g, u) = \{(0, 0)$, $(0, 3)$, $(3, 0)$, $(3, 3)$, $(1.45, 0)$, $(1.45, 3)\}$. The order parameter has converged within acceptable numerical precision by $\tau = 0.0375$ $(2\pi B)^{-1}$. Similar behavior is observed for the total energy, and for the out-of-plane polarization, $\phi_z$.

Analogously, Figure~\ref{fig:square_tau} shows the behavior of $\phi_{xy}$ as a function of $\tau$ at this same value of $\beta$ and with $(g, u) = \{(0, 0)$, $(0, 3)$, $(3, 0)$, $(3, 3)$, $(1.25, 0)$, $(1.25, 3)\}$ for dipoles on a square lattice with 64 dipoles in the periodic simulation cell. As with the triangular lattice simulations, the variation of $\phi_{xy}$ is converged within acceptable numerical precision by $\tau = 0.0375$ $(2\pi B)^{-1}$. Just as for the triangular lattice, similar behavior is also observed for the total energy and for $\phi_z$.

Having established acceptable values for the short time step, $\tau = 0.0375$ $(2\pi B)^{-1}$, balancing accuracy and efficiency, it is then necessary to establish the required path length in imaginary time to ensure sampling of the ground state, and to make sure that the trial wave function, Eq.~(\ref{eq:approx_wf}), is not biasing the results. Figure~\ref{fig:square_beta} shows the behavior of $\phi_{xy}$ as a function of $\beta$ for $(g, u) = \{(0, 0)$, $(0, 3)$, $(3, 0)$, $(3, 3)$, $(1.25, 0)$, $(1.25, 3)\}$. For $g = 0$ and $3$ it appears that the quantities of interest are converged within acceptable tolerances by $\beta = 4.2$ $(2\pi B)^{-1}$. What is interesting is the behavior of $\phi_{xy}$ with respect to $\beta$ at $g = 1.25$. While the decay of $\phi_{xy}$ with respect to $\beta$ is very rapid at $g = 0$ and $3$, at $g = 1.25$ the convergence is very slow. This slow convergence with respect to $\beta$ was not observed for the energy.

For a better idea of how widespread this slow convergence is, a series of simulations at $u = 0.5$ and various values of $g$ were undertaken with 48 dipoles on a triangular lattice. Figure~\ref{fig:triangular_beta} shows the behavior of $\phi_\mathrm{pol}$ as a function of $\beta$ and $g$. It appears that for values of $g \leq 1.25$ and $g \geq 1.75$, $\phi_\mathrm{pol}$ is effectively converged well before $\beta = 5.1$ $(2 \pi B)^{-1}$. However, for $g = 1.5$ the value of $\phi_\mathrm{pol}$ converges very slowly. One plausible explanation for this is the presence of a second order phase transition, one manifestation of which would be a diverging spatial correlation length at the transition. In this case the Hartree trial wave function, Eq.~(\ref{eq:approx_wf}), which takes the form of a product of single particle functions, would be qualitatively incorrect. As a result it would have very poor overlap with the true ground state wave function, and so the decay of relevant quantities toward their ground state values would expected to be very slow.
In addition, a second order phase transition is characterized by soft modes
(Goldstone modes) with a vanishing excitation energy as we approach the phase transition.
This is the reason for the slow dynamics near the second order phase transition. 
In the imaginary time evolution used in PIGS, such soft modes decay with
a large time constant that, in the thermodynamic limit, diverges at the quantum phase transition,
leading to a slow convergence to the ground state as a function
of the imaginary time path length $\beta$. Note that due to the small excitation energy of a soft mode, the total energy is barely affected, which is consistent with its observed rapid convergence with
$\beta$.  For this reasons, we propose that the quantum phase transitions studied in this work are of second order.

Although the goal of quantum Monte Carlo simulations is a quantiative description,
the convergence study shows that close to the quantum phase transition to the
orientationally ordered phase, our simulations are still biased
by the value of the imaginary time path length $\beta$.
From Figures~\ref{fig:square_beta} and \ref{fig:triangular_beta}
we see that at the critical $g$, $\beta$ would need to be much larger than the value $\beta = 4.2$ $(2 \pi B)^{-1}$ and $\beta = 5.1$ $(2 \pi B)^{-1}$ used in our simulations of square and triangular lattices, respectively. In fact, in the thermodynamic limit $\beta$ would need to be infinitely large for
the critical value of $g$.  Determining the exact values of the order
parameters  $\phi_\mathrm{pol}$ and $\phi_{xy}$ very close to the critical value of $g$ requires an extensive (and computationally
expensive) suite of simulations with increasing $\beta$
and careful extrapolations to $\beta\to\infty$, in addition to undertaking a finite size scaling analysis
typically used in studies of second order phase transition.

The most efficient solution to reduce the bias is by reducing the excited state contribution
to the trial wave function.  Rather than relying on a Hartree trial wave function Eq.~(\ref{eq:approx_wf}),
correlated trial wave functions need to be designed and optimized.  For example, an optimized Jastrow ansatz
would include pair correlations.  This goes beyond the scope of this paper but we note that the
hypernetted-chain Euler-Lagrange method adapted to orientational degrees of freedoms\cite{hufnagldiss}
could provide a highly optimized Jastrow ansatz, with presumably very small
overlap with excited states, such that even small $\beta$ values give accurate results.

\end{appendix}


\end{document}